\documentclass[final,1p,times, authoryear]{elsarticle}

\makeatletter
\def\ps@pprintTitle{%
 \let\@oddhead\@empty
 \let\@evenhead\@empty
 \def\@oddfoot{\centerline{\thepage}}%
 \let\@evenfoot\@oddfoot}
\makeatother

\usepackage{amssymb}
\usepackage{amsthm}
\usepackage[tbtags, sumlimits, intlimits]{amsmath}
\usepackage{tensor}
\usepackage{empheq}
\usepackage{pdflscape}
\usepackage{multirow}
\usepackage{siunitx}
\usepackage{subfigure}
\usepackage{setspace} 
\usepackage{mathrsfs}
\usepackage{textcomp}
\usepackage{color}

\biboptions{sort}
\usepackage{graphicx,psfrag,epsfig,rotating}
\usepackage{color}

\usepackage{booktabs}
\usepackage{url}
\usepackage{longtable}

\usepackage{multirow}
\usepackage{setspace}
\usepackage{nomencl}
\usepackage{lscape}
\usepackage{tabularx}
\usepackage{pdflscape}

\usepackage{array,float,latexsym,amsmath,amssymb,amsbsy,eurosym,theorem,amsfonts,mathrsfs,bbm,verbatim}

\usepackage[active]{srcltx}

\usepackage{pdfsync}

\newcommand{\dfracp}[2]{\dfrac{\partial #1}{\partial#2}}
\newcommand{\jmp}[1]{[\![ {#1} ]\!]}

\newcommand{\eqn}[1]{Eq.~(\ref{#1})}

\newcommand{\fig}[1]{Fig.~\ref{#1}}

\newcommand{\sect}[1]{Sec.~\ref{#1}}

\renewcommand{\b}[1]{\boldsymbol{#1}} 

\newcommand{\h}[1]{\widehat{#1}}
\renewcommand{\o}[1]{\overline{#1}}
\renewcommand{\u}[1]{\underline{#1}}
\renewcommand{\t}[1]{\widetilde{#1}}
\renewcommand{\u}[1]{\underline{#1}}

\newcommand{\lrb}[1]{\left[ {#1} \right]}
\newcommand{\lrrb}[1]{\left( {#1} \right)}

\newcommand{\sfrac}[2]{\mbox{$\frac{#1}{#2}$}}
\newcommand{\infrac}[2]{{#1}/{#2}}

\newcommand{\grad}{\nabla}

\renewcommand{\div}{\mbox{div}}
\newcommand{\curl}{\mbox{curl}}

\newcommand{\dyad}{\otimes}

\newcommand{\prs}{{}^{\mbox{\scriptsize p}}} 

\newcommand{\trns}{{}^{\mathsf{\scriptsize T}}}

\newcommand{\sth}{\{\bullet\}}

\definecolor{gray}{rgb}{0.75, 0.75, 0.75}
\definecolor{yellow}{rgb}{1, 0.7, 0.2}
\definecolor{green}{rgb}{0.3, 0.9, 0.3}
\definecolor{brown}{rgb}{0.6, 0.3, 0.2}
\definecolor{magenta}{rgb}{0.9, 0.1, 0.9}
\definecolor{light}{rgb}{1, 0.7, 0.7}

\journal{International Journal of Plasticity}

\newproof{pot}{Remark}
\newproof{prop}{Proposition}
\newproof{a_proof}{Proof}

\begin{document}

\begin{frontmatter}



\title{Computational and theoretical aspects of a grain-boundary model that accounts for grain misorientation and grain-boundary orientation }

%
%
\author[icm]{D. Gottschalk}
\ead{gottschalk@ikm.uni-hannover.de} 

\author[uct]{A. McBride\corref{cor1}}
\ead{andrew.mcbride@uct.ac.za}

\author[uct]{B.D. Reddy}
\ead{daya.reddy@uct.ac.za}

\author[erlangen]{A. Javili}
\ead{ali.javili@ltm.uni-erlangen.de}

\author[icm]{P. Wriggers}
\ead{wriggers@ikm.uni-hannover.de}

\author[icm]{C.B. Hirschberger}
\ead{c.b.hirschberger@gmail.com}

\address[icm]{
Institute of Continuum Mechanics,
Leibniz Universit{\"a}t Hannover,
Appelstrasse 11,
30167 Hannover,
Germany
}  
 
\address[uct]{
Centre for Research in Computational and Applied Mechanics,
University of Cape Town,
5th floor, Menzies Building,
Private Bag X3,
7701 Rondebosch,
South Africa
Tel: + 27 21 650-3817
Fax: + 27 21 685-2281}

\address[erlangen]{
Chair of Applied Mechanics, 
University of Erlangen--Nuremberg, 
Egerlandstr. 5, 
91058 Erlangen,
Germany, 
Tel.: +49 (0)9131 85 28502, 
Fax: +49 (0)9131 85 28503}

\cortext[cor1]{Corresponding author}

\begin{abstract}

A detailed theoretical and numerical investigation of the infinitesimal single-crystal gradient-plasticity and grain-boundary theory of \citeauthor{Gurtin2008} (2008) ``A theory of grain boundaries that accounts automatically for grain misorientation and grain-boundary orientation". {\it Journal of the Mechanics and Physics of Solids} {\bf 56} (2), 640--662, is performed. 
The governing equations and flow laws are recast in variational form. 
The associated incremental problem is formulated in minimization form and provides the basis for the subsequent finite element formulation. 
Various choices of the kinematic measure used to characterize the ability of the grain boundary to impede the flow of dislocations are compared. 
An alternative measure is also suggested. 
A series of three-dimensional numerical examples serve to elucidate the theory. 
\end{abstract}


\end{frontmatter}

\newpage

\section{Introduction}\label{sec_intro}

The miniaturisation of mechanical components composed of crystalline material requires a continuum theory that accounts for the role of the grain boundary and for size-dependent effects.
The grain-boundary model should incorporate both the \emph{misorientation in the crystal lattice} between adjacent grains, and the \emph{orientation of the grain boundary} relative to the crystal lattice of the adjacent grains.
Classical theories of plasticity are unable to describe the well-known size-dependent response exhibited by crystalline material at the micro- and nanometre scale. 
Numerous extended (gradient and non-local) continuum theories of single-crystal plasticity have been presented in the last two decades to circumvent these limitations. 
The thermodynamically consistent gradient theory of \citeauthor{Gurtin2008b} and co-workers and related works \citep[see e.g.][]{Gurtin2002, Gurtin2005, Gurtin2006, Gurtin2008b} have received particular attention.
A variational formulation of the \citet{Gurtin2002} framework has been developed in \citet{Reddy2011a, Reddy2011}. 
In \citet{Gurtin2002} the defect part of the free energy is parametrized in terms of the (bulk) Burgers tensor, a rigorously defined and physically meaningful measure of the (local) Burgers vector and hence the lattice mismatch \citep[see e.g.][]{Nye1953}.
The form of the defect energy was modified by \citet{Gurtin2008b} to account for a continuous distribution of geometrically necessary dislocations (GNDs). 
The recent work of \citet{Gurtin2014} uses a scalar measure of the accumulated slips as the basis for the  hardening relation, which takes account of both self- and latent-hardening. 
Furthermore, the resulting initial boundary-value problem is placed in a variational setting in the form of a global variational inequality.
\cite{Erturk2009} show how the theory of \citeauthor{Gurtin2002} et al.\ can be related to the more physically motivated theories due to \citet{Evers2002, Evers2004, Evers2004a, Bayley2006}.

The gradient theory of \citet{Gurtin2008b} provides a basis to account for the role of the grain boundary \citep[see][]{Gurtin2008}. 
Neumann and Dirichlet-type boundary conditions on the slip and the flux of the vectorial microforce (i.e.\ the microscopic traction), respectively, can be prescribed and are often assumed homogeneous. The homogeneous Dirichlet condition, known as the micro-hard boundary condition, has been widely used to account for the grain boundary or an interface \citep[see e.g.][]{Evers2004, Ekh2007, Kuroda2008, Lele2008, Ohno2007}. 
Clearly this boundary condition ignores the complex geometric structures in the vicinity of the grain boundary.

Central to the theory of \citet{Gurtin2008} is the introduction of the \emph{grain-boundary Burgers tensor} to parametrize the grain-boundary free energy. 
The grain-boundary Burgers tensor is obtained from the mismatch in the plastic part of the displacement gradient around a circuit centred on the grain boundary and contains information on both the misorientation in the crystal lattice between adjacent grains, and the orientation of the grain boundary. 
Furthermore, the definition of grain-boundary Burgers tensor is consistent with that of the Burgers tensor in the bulk.  

The grain-boundary Burgers tensor can be expressed in terms of the \emph{intra}- and \emph{inter-grain interaction moduli}. 
The inter-grain interaction moduli account for mismatch in the slip systems adjacent to the grain boundary and the orientation of the grain boundary.
They provide a physically meaningful characterisation of the interaction of neighbouring slip systems with the extremes described as \emph{non-interactive} and \emph{maximally interactive}.

Recently, \citet{Beers2013} have proposed and numerically implemented a theory similar to that of \citet{Gurtin2008} for incorporating grain boundaries into the gradient crystal plasticity theory of \citet{Evers2004}.  
A key feature of the theory is the use of a geometrically-motivated vectorial measure to parametrize the grain-boundary free energy. 
We will show in this work that under \emph{planar conditions} (as investigated numerically by \citet{Beers2013}) the models of \citet{Gurtin2008} and \citet{Beers2013} produce identical interaction moduli.  
In similar work, \citet{Oezdemir2014} implemented the grain-boundary theory of \citet{Gurtin2008}.
A series of finite element simulations of planar bi-crystals (single and double slip) illustrated features of the grain-boundary model.

\citet{Gudmundson2004} and \citet{Fredriksson2005, Fredriksson2005a} propose an interface theory in which both the interface moment traction and the plastic slip can be discontinuous at the grain boundary.
The model introduces an interfacial free energy that depends on the plastic strain  on both sides of the interface. 
Critically however, the model does not account directly for the mismatch in the adjacent grains or the orientation of the grain boundary.
Related works include those by \citet{Aifantis2005, Aifantis2006}.
\cite{Ekh2011} propose a ``micro-flexible'' grain boundary which provides a degree of resistance to plastic flow dependent upon the misorientation of the adjacent grains \citep[see][for an extension that accounts for thermal effects]{Bargmann2013}. 
They do not consider the orientation of the grain boundary.
\citet{Wulfinghoff2013} account for grain boundaries within a gradient-plasticity theory by postulating a grain-boundary yield condition and flow rule.
The theory does not account for the mismatch in the adjacent grains or the orientation of the grain boundary.
\citet{Voyiadjis2014} developed and numerically implemented a gradient-plasticity model for the polycrystalline problem which accounts for the role of grain boundaries via the mismatch in the accumulated plastic strain.

Recasting the problem of single-crystal gradient plasticity as a variational formulation makes it amenable to analysis \citep[see][]{Reddy2011}. 
The variational formulation does not have an associated minimization problem, but the corresponding time-discrete incremental problem does. 
The variational formulation developed in \citet{Reddy2011} is extended here to include the grain boundary and a viscoplastic flow law. 
The associated incremental minimization problem is shown to be equivalent to the time-discrete variational problem and provides the point of departure for the numerical implementation within the finite element method. 
The software AceGen \citep{Korelc2002} is used to describe the finite element interpolation,  and to compute the residual and (algorithmically consistent) tangent contributions directly from the prescribed functional associated with the incremental minimization problem, using automatic differentiation at the level of the quadrature point. 
This approach ensures quadratic convergence of the algorithm and greatly simplifies the implementation. 
Details of the numerical implementation are given. 

\citet{Gurtin2008} proposes two thermodynamically admissible plastic flow relations for the grain boundary (denoted \citeauthor{Gurtin2008}~I and II). 
The flow relations define the structure of the dissipative microscopic stress in the grain boundary microscopic force balance. 
The flux of dislocations from the grains drives the microscopic force balance. 
In the first proposal, the grain boundary Burgers tensor is used to parametrize the flow relations, while in the second it is the slip. 
The first approach accounts for the interaction of slip systems adjacent to the grain boundary.
This approach also allows for a recombination of the plastic distortion contributions from adjacent sides via the definition of the grain boundary Burgers tensor. 
The second approach does not directly account for the structure of the adjacent grains or the orientation of the grain boundary in the plastic flow relation. 
Both approaches account for the geometric structure of the adjacent grains and the grain boundary via the flux terms from the grains.

A series of three-dimensional numerical examples  elucidate the grain-boundary theory. 
The examples demonstrate single slip in a bi-crystal and multi-slip in a polycrystal where each of the 27 grains is a face-centered-cubic crystal structure. 
The polycrystal example in particular demonstrates various features of the  \citet{Gurtin2008} theory that are not obvious from the theory or the single slip examples. 
The \citeauthor{Gurtin2008}~I model for the plastic flow relation is unable to capture the widely-used micro-hard condition in multi-slip problems, even when using an artificially high value for grain boundary slip resistance. 
The \citeauthor{Gurtin2008}~II model can capture the range of responses between the micro-free and micro-hard conditions. 
Motivated by a plastic flow relation that accounts for the structure of the grain boundary and captures the micro-hard and micro-free limits and the range between, a modified measure of the grain-boundary Burgers tensor is analysed and implemented. 
In the modified formulation, the micro-hard limit is recovered for large-angle grain boundaries, and the micro-free for perfectly aligned crystal structures on either side of the grain boundary.  

%

The structure of this work is as follows. 
The kinematics of the gradient crystal plasticity formulation in the bulk and on the grain boundary  are reviewed in \sect{sec_kinematics}. 
The kinematic measures of the mismatch at the grain boundary proposed in \citet{Gurtin2008} and \cite{Beers2013} are then compared. 
Particular attention is paid to the inter-grain interaction moduli.
The kinetics of the problem and the various restrictions to the theory assumed are presented in \sect{sec_kinetics}. 
The governing relations and the plastic flow law are presented in \sect{sec_governing_equations} and \ref{sec_flow_relations}.
An alternative measure of the kinematic mismatch at the grain boundary is given.
The variational formulation of the problem and the associated incremental formulation are developed in \sect{sec_variational_problem}. 
This is followed by details of the numerical implementation within the finite element framework. 
The finite element model is then used to simulate a series of representative numerical examples in \sect{sec_numerical_examples}. 
Finally, conclusions are made and various extensions proposed. 

\subsection*{Notation and basic relations}

Direct notation is adopted throughout. 
Occasional use is made of index notation, the summation convention for repeated indices being implied. 
When the repeated indices are lower-case italic letters, the summation is over the range $\{1,2,3\}$.
Upper-case italic indices can refer to arbitrary adjacent grains $\{ A,~B\}$.
The summation convention is not employed for grains.
The scalar product of two vectors $\b{a}$ and $\b{b}$ is denoted $\b{a}\cdot\b{b} = [\b{a}]_{i} [\b{b}]_{i}$.  
The scalar product of two second-order tensors $\b{A}$ and $\b{B}$ is denoted $\b{A}:\b{B} = [\b{A}]_{ij} [\b{B}]_{ij}$. 
The composition  of two second-order tensors $\b{A}$ and $\b{B}$, denoted $\b{A}  \b{B}$, is a second-order tensor with components  $[\b{A} \b{B}]_{ij} = [\b{A}]_{im} [\b{B}]_{mj}$.
The tensor product of two vectors $\b{a}$ and $\b{b}$ is a second-order tensor $\b{D}=\b{a}\dyad\b{b}$ with $[\b{D}]_{ij}= [\b{a}]_{i} [\b{b}]_{j}$. 
The action of a second-order tensor $\b{A}$ on a vector $\b{b}$ is a vector with components $[\b{a}]_i = [\b{A}]_{im} [\b{b}]_m$. 
The curl of a second-order tensor $\b{A}$ is a second-order tensor with components $[\curl \b{A}]_{ij} = \epsilon_{irs} \partial A_{js} / \partial x_r$, where $\b{\epsilon}$ is the third-order permutation tensor.
An arbitrary quantity in the bulk is denoted $\sth$ and analogously $\{ \o{\bullet} \}$ denotes an arbitrary quantity on the grain boundary.
Any array associated with the set of $N$ slip systems is denoted
$
\u{\gamma}:= \{ \gamma^1 \, , \,  \gamma^2  \, \ldots \, , \gamma^N \} \, .
$
Summation over the slip systems will be abbreviated by $\sum_\alpha$. 
Index notation is not employed for summations over slip systems.
The unit basis vectors in the Cartesian (standard-orthonormal) basis are $\{ \b{e}_1,~\b{e}_2,~\b{e}_3 \}$.
The following identities are employed widely:
\begin{align*}
	[\b{n} \times]_{ij} &:= \epsilon_{ikj}n_k \, ,\\
	  -[\b{n} \times][\b{n} \times] &= \b{I} - \b{n} \otimes \b{n} \, ,\\
 [\b{n} \times]  \b{b} &= \b{n} \times \b{b} \, ,
\end{align*}
where $\b{I}$ is the second-order identity tensor.

\section{Kinematics}\label{sec_kinematics}

The kinematic basis of the gradient-plasticity and grain-boundary theory of \citet{Gurtin2008} is first summarised. 
A comparison of aspects of the theory with that of \citet{Beers2013} is then made. 
The theory presented is based on the assumption of infinitesimal deformations. 
The initial configuration is thus assumed to be geometrically representative for all time.

\subsection{Bulk}

\begin{figure}[!ht]
 \centering
 \includegraphics[width = 0.35\textwidth]{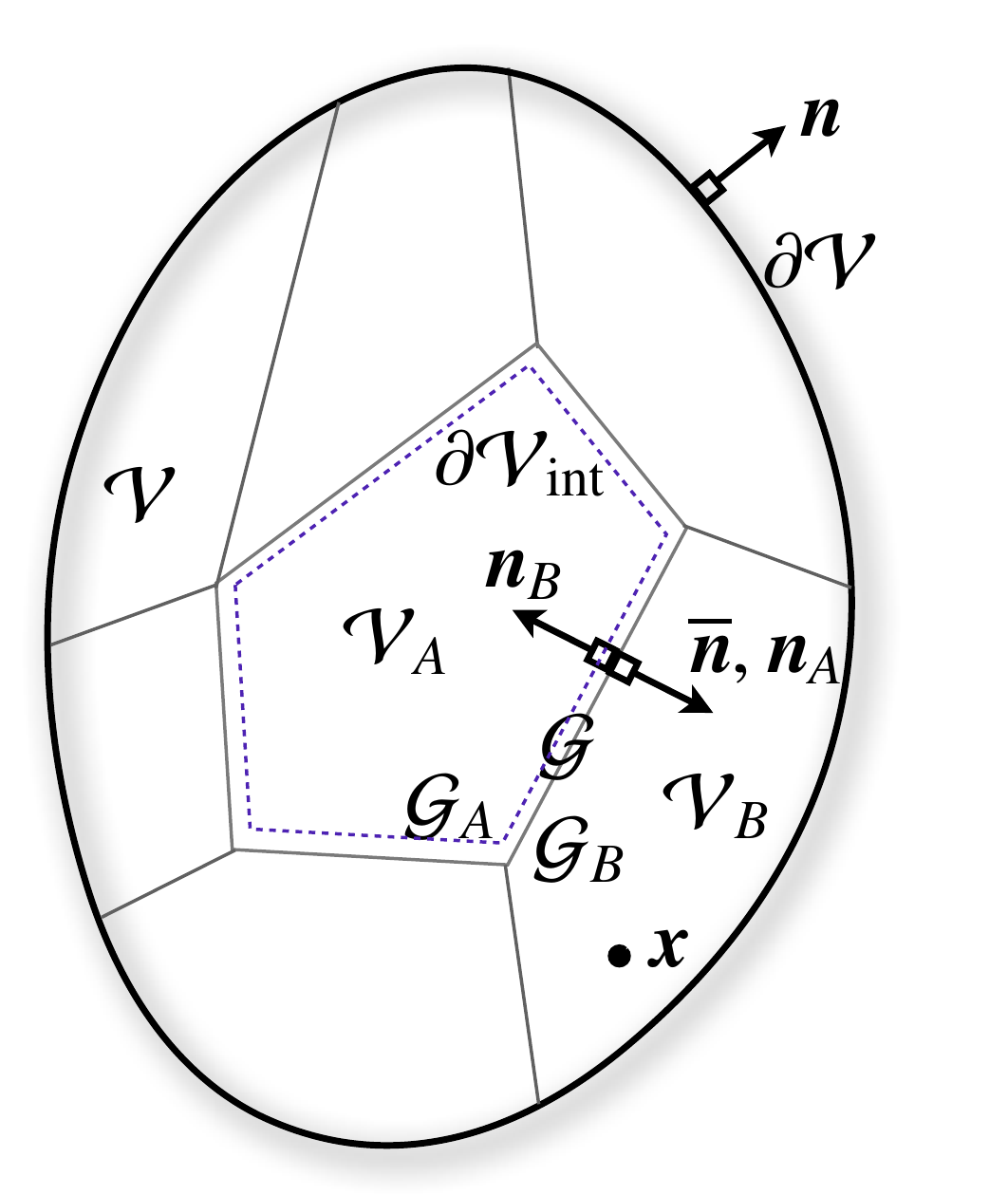}
 \caption{A continuum body $\mathcal{V}$, composed of grains $\mathcal{V}_I$ separated by a two-sided grain boundary $\mathcal{G}$.} 
 \label{fig_domains}
\end{figure}

Consider a continuum body whose placement $\mathcal{V}$ at time $t=0$ is shown in \fig{fig_domains}. 
A  typical material point is identified by the position vector $\b{x}$. 
The displacement of the material point is denoted by $\b{u}(\b{x},t)$. 
The displacement gradient $\b{H} := \nabla \b{u}$ is decomposed (locally) into elastic and  plastic parts $\b{H}^\text{e}$ and $\b{H}^\text{p}$ according to
\begin{gather*}
\b{H} = \b{H}^\text{e} + \b{H}^\text{p} \, .
\end{gather*}
The elastic displacement gradient  $\b{H}^\text{e}$ accounts for recoverable elastic lattice stretching, while $\b{H}^\text{p}$ quantifies the plastic distortion due to slip on the predefined slip planes. 
The elastic strain $\b{E}^\text{e}$ is given by the symmetric part of the elastic displacement gradient
\begin{align*}
	\b{E}^\text{e} := \sfrac{1}{2} \lrb{ \b{H}^\text{e} + \b{H}^\text{e}{}\trns }  \, .
\end{align*}
The flow of dislocations through the crystal lattice is described kinematically via the assumption that the plastic distortion tensor can be expressed in terms of the slip $\gamma^\alpha$ on the individual prescribed slip systems $\alpha = 1,2,\ldots,N$ as
\begin{align}
\b{H}^\text{p} = \sum_\alpha \gamma^\alpha \b{s}^\alpha  \otimes \b{m}^\alpha 
	= \sum_\alpha \gamma^\alpha \mathbb{S}^\alpha \, . \label{H_p}
\end{align}
The slip direction and slip plane normal of slip system $\alpha$ are denoted $\b{s}^\alpha$ and $\b{m}^\alpha$, respectively, where $\b{s}^\alpha \cdot \b{m}^\alpha = 0$ and $ \vert \b{s}^\alpha \vert = \vert \b{m}^\alpha \vert = 1$. 
The vector $\b{l}^\alpha$  is defined by $\b{l}^\alpha := \b{m}^\alpha \times \b{s}^\alpha$. 
Hence $\{ \b{m}^\alpha, \b{s}^\alpha, \b{l}^\alpha \}$ constitute a local orthonormal basis. 
The Schmid tensor is defined by $\mathbb{S}^\alpha = \b{s}^\alpha \otimes \b{m}^\alpha$. 
The Burgers tensor
\begin{align}
	\b{G} =  \curl \b{H}^\text{p} 
	= \sum_\alpha \lrb{\grad \gamma^\alpha \times \b{m}^\alpha}\otimes \b{s}^\alpha \, , \label{G_bulk}
\end{align}  
quantifies the crystal distortion due to dislocations. 
It is obtained from the boundary integral of $\b{H}^\text{p}$ over an infinitesimal closed circuit in the bulk.
The vector $\b{G}\trns \b{e}$ gives the Burgers vector, per unit area, for a closed circuit on a plane with unit normal $\b{e}$.

Following \citet{Gurtin2008b}, the constitutive theory at the microscopic scale accounts for a continuous distribution of GNDs. 
The dislocations are either of edge or screw type and are characterised in terms of their Burgers and line directions as follows:
\begin{itemize}
\item \emph{edge dislocation}: Burgers direction $\b{s}^\alpha$ and line direction $\b{l}^\alpha$;
\item \emph{screw dislocations}:  Burgers direction $\b{s}^\alpha$ and line direction $\b{s}^\alpha$.
\end{itemize}
The density of the edge and screw  dislocations per unit area are related to the slip gradient as
\begin{align}
 \rho^\alpha_\vdash = -  \frac{1}{b}\grad \gamma ^\alpha \cdot \b{s}^\alpha
&& \text{and} &&
 \rho^\alpha_\odot = \frac{1}{b} \grad \gamma^\alpha \cdot \b{l}^\alpha \, ,
 \label{edge_screw}
\end{align}
where $b$ is the length of the Burgers vector. 

\subsection{Grain boundary}

The body $\mathcal{V}$ is composed of grains separated by grain boundaries as depicted in \fig{fig_domains}.
The boundary of a grain interior to the domain is denoted by $\partial \mathcal{V}_\text{int}$.
The set of all grains is denoted by $\mathsf{V} = \{ \mathcal{V}_A \}$, with $\mathcal{V}_A$ an arbitrary grain. 
The grain boundary is represented as a two-sided interface, with common sides $\mathcal{G}_A$ and $\mathcal{G}_B$. 
The set of all grain boundaries is denoted by $\mathsf{G}=\{ \mathcal{G}_C \}$, where $C$ runs from one to the number of grain boundaries.
The grains on either side of the grain boundary are correspondingly denoted by $\mathcal{V}_A$ and $\mathcal{V}_B$. 
The outward unit normals to $\mathcal{G}_A$ and $\mathcal{G}_B$ are denoted $\b{n}_A$ and $\b{n}_B$, respectively. 
The normal to the grain boundary  is defined by $\o{\b{n}} := \b{n}_A$.  
The displacement field across the grain boundary is assumed to be \emph{continuous} \citep[see][for a model of grain boundaries that slip and separate]{Gurtin2008a}; that is,
\begin{align}
\jmp{\b{u}} := \b{u}\rvert_{\mathcal{G}_B} - \b{u}\rvert_{\mathcal{G}_A}   = \b{0} 
   && \text{and} && 
   \o{\b{u}} &:= \b{u} \rvert_{\mathcal{G}} 
   \, . \label{coherence}
\end{align}
The plastic slip  and its gradient are however generally discontinuous across the grain boundary.
%
Central to the grain-boundary theory of \citeauthor{Gurtin2008} is the \emph{grain-boundary Burgers tensor}
\begin{align*}
	\o{\b{G}} = \jmp{\b{H}\prs}  [\o{\b{n}} \times] =: \o{\b{H}}\prs [\o{\b{n}} \times]  \, , 
\end{align*}
obtained by integrating the plastic distortion $\b{H}^\text{p}$ along an infinitesimal circuit centred on the grain boundary (cf.\ \eqn{G_bulk}). 
The grain-boundary Burgers tensor can be expressed in terms of the slip on either side of the grain boundary, the orientation of the slip systems and the grain-boundary orientation as
\begin{align*}
	\o{\b{G}}
		&=  \sum_\alpha \lrb{ \gamma^\alpha_B {\o{\mathbb{N}}{}^\alpha_B} 
		- \gamma^\alpha_A  {\o{\mathbb{N}}{}^\alpha_A}} 
		= \sum_\alpha \o{\b{G}}{}^\alpha
		  \, , 
\end{align*}
where
\begin{align*}
\o{\mathbb{N}}{}_J^\alpha = \mathbb{S}_J^\alpha [\o{\b{n}} \times] 
	= \b{s}^\alpha_J \otimes \lrb{\b{m}^\alpha_J \times \o{\b{n}}} && \text{for } J=A,B \, ,
\end{align*}
is the Schmid orientation tensor for system $\alpha$ in grain $J$. 

The Burgers vector production within the grain boundary is given by 
\begin{align}
	\dot{\o{\b{G}}} &=  \dot{\o{\b{H}}}\prs [\o{\b{n}} \times] = 
	\sum_\alpha \lrb{ \dot{\gamma}{}^\alpha_B {\o{\mathbb{N}}{}^\alpha_B} 
		- \dot{\gamma}{}^\alpha_A  {\o{\mathbb{N}}{}^\alpha_A}}
	=  \sum_\alpha \dot{\o{\b{G}}}{}^\alpha	 \, . \label{G_bar_dot}
\end{align}

The contraction of the Schmid orientation tensor  with the grain-boundary Burgers tensor motivates the definition of  the slip-interaction moduli $\o{\mathbb{C}}{}_{JB}^{\alpha \beta} := \o{\mathbb{N}}{}_J^\alpha : \o{\mathbb{N}}{}_B^\beta$, that is,
\begin{gather*}
\begin{split}
\o{\mathbb{N}}{}_J^\alpha : \o{\b{G}}
	&=  \sum_\beta \lrb{ \gamma^\alpha_B  \lrb{ \o{\mathbb{N}}{}_J^\alpha : \o{\mathbb{N}}{}_B^\beta} 
		- \gamma^\alpha_A  \lrb{\o{\mathbb{N}}{}_J^\alpha : \o{\mathbb{N}}{}^\beta_A}} \\
	&= 	\sum_\beta  \lrb{ \gamma^\alpha_B \o{\mathbb{C}}{}_{JB}^{\alpha \beta} - 
		\gamma^\alpha_A \o{\mathbb{C}}{}_{JA}^{\alpha \beta}} \, , 
\end{split}		
\end{gather*}
where
\begin{align}
\o{\mathbb{C}}{}_{AB}^{\alpha \beta} = \o{\mathbb{C}}{}_{BA}^{\beta \alpha} \, , && 
\o{\mathbb{C}}{}_{AA}^{\alpha \beta} = \o{\mathbb{C}}{}_{AA}^{\beta \alpha} \, , && 
\o{\mathbb{C}}{}_{BB}^{\alpha \beta} = \o{\mathbb{C}}{}_{BB}^{\beta \alpha} \, .\label{interaction_sym}
\end{align} 
Furthermore,
\begin{align}
\o{\mathbb{C}}{}_{IJ}^{\alpha \beta} = [\b{s}^\alpha_I \cdot \b{s}^\beta_J][\b{m}^\alpha_I \times \o{\b{n}}] \cdot [\b{m}^\beta_J \times \o{\b{n}}] \, . \label{C_AB_Gurtin}
\end{align}
The slip interaction moduli are classified as:
\begin{itemize}
\item $\o{\mathbb{C}}{}_{AA}^{\alpha \beta}$: \emph{intra-grain} interaction moduli for grain A;
\item $\o{\mathbb{C}}{}_{AB}^{\alpha \beta}$: \emph{inter-grain} interaction moduli between grains A and B;
\item $\o{\mathbb{C}}{}_{BB}^{\alpha \beta}$: \emph{intra-grain} interaction moduli for grain B.
\end{itemize}
The inter-grain interaction moduli for an arbitrary pair of slip systems and grain-boundary orientation thus have the range $-1 \leq  \mathbb{C}_{AB}^{\alpha \beta} \leq 1$.
The inter-grain interaction moduli characterise the interaction between slip systems in adjacent grains separated by a grain boundary. 
The systems are \emph{non-interactive} if $\mathbb{C}_{AB}^{\alpha \beta} = 0$ and \emph{maximally interactive} if $\vert \mathbb{C}_{AB}^{\alpha \beta} \vert = 1$. 
These definitions imply that a pair of slip systems in adjacent grains are (see \eqn{C_AB_Gurtin}):
\begin{itemize}
\item \emph{non-interactive} if and only if either:
	\begin{align*}
		\b{s}_A^\alpha \cdot \b{s}_B^\beta = 0
		&& \text{or} && 
		[\b{m}_A^\alpha \times \o{\b{n}}] \cdot [\b{m}_B^\beta \times \o{\b{n}}] = 0 \, ;
	\end{align*}
\item \emph{maximally interactive} if and only if all of the following  conditions are satisfied:
	\begin{align*}
	\b{s}_A^\alpha = \pm \b{s}_B^\alpha \, , &&
	\b{m}_A^\alpha \cdot \o{\b{n}} = 0 \, , &&
	\b{m}_B^\beta \cdot \o{\b{n}} = 0 \, .
	\end{align*}
\end{itemize}

A key measure in the \citet{Gurtin2008} model used to parametrize the grain-boundary free energy is the norm of the grain-boundary Burgers tensor:
\begin{align}
\bigl| \o{\b{G}} \bigr|^2 &:= \o{\b{G}} : \o{\b{G}} 
= \bigl| \sum_\alpha \o{\b{G}}{}^\alpha \bigr|^2
=
	\sum_{\alpha,\beta}\biggl[
	\gamma^\alpha_B \gamma^\beta_B \underbrace{\o{\mathbbm{N}}{}^\alpha_B : \o{\mathbbm{N}}{}^\beta_B}_{\o{\mathbb{C}}{}^{\alpha \beta}_{BB}}
	- 2 \gamma^\alpha_A \gamma^\beta_B \underbrace{\o{\mathbbm{N}}{}^\alpha_A : \o{\mathbbm{N}}{}^\beta_B}_{\o{\mathbb{C}}{}^{\alpha \beta}_{AB}}
	+ \gamma^\alpha_A \gamma^\beta_A \underbrace{\o{\mathbbm{N}}{}^\alpha_A : \o{\mathbbm{N}}{}^\beta_A}_{\o{\mathbb{C}}{}^{\alpha \beta}_{AA}}\biggr] \, , \label{G_norm_gurtin}
\end{align}
which is a function of the adjacent slip system structures and the grain-boundary orientation.

\subsection{Comparison of aspects of the \citeauthor{Gurtin2008} and \citeauthor{Beers2013} models}\label{sec_comp_gurtin_beers}

The kinematic measures of the mismatch at the grain boundary employed in the models of \cite{Beers2013} and \cite{Gurtin2008} are now compared. 
The model of \cite{Beers2013} is stated using notation similar to \citet{Gurtin2008} to aid comparison.

\Citet{Beers2013} define the geometrically-motivated grain-boundary normal slip components \citep[see][]{Kuroda2008, Erturk2009}  by
\begin{align*}
q^\alpha_A := \gamma^\alpha_A \lrb{ \b{s}^\alpha_A - \b{l}^\alpha_A } \cdot \o{\b{n}} 
&& \text{and} &&
q^\alpha_B := \gamma^\alpha_B \lrb{ -\b{s}^\alpha_B + \b{l}^\alpha_B } \cdot \o{\b{n}} \, .
\end{align*} 
The net defect vector (modelling parameters are set to unity)  on the grain boundary is defined by
\begin{align*}
\o{\b{g}} 	&:= \sum_\alpha \lrb{q^\alpha_A \b{s}^\alpha_A + q^\alpha_B \b{s}^\alpha_B} \, , \notag
\intertext{which can be restated in a form conducive for comparison as}
\o{\b{g}}			&= \sum_\alpha \biggl[ \gamma^\alpha_B\underbrace{\lrb{\b{s}^\alpha_B \otimes \b{l}^\alpha_B - \b{s}^\alpha_B \otimes \b{s}^\alpha_B }  \o{\b{n}}}_{\o{\mathbbm{n}}{}^\alpha_B}
			-\gamma^\alpha_A\underbrace{ \lrb{\b{s}^\alpha_A \otimes \b{l}^\alpha_A - \b{s}^\alpha_A \otimes \b{s}^\alpha_A}  \o{\b{n}}}_{\o{\mathbbm{n}}{}^\alpha_A} 
			 \biggr] \, , \notag  \\ 
\implies \quad \vert \o{\b{g}} \vert^2 &=
	\sum_{\alpha,\beta} \biggl[
	\gamma^\alpha_B \gamma^\beta_B \underbrace{\o{\mathbbm{n}}{}^\alpha_B \cdot \o{\mathbbm{n}}{}^\beta_B}_{\o{\mathbbm{c}}{}^{\alpha \beta}_{BB}}
	- 2 \gamma^\alpha_A \gamma^\beta_B \underbrace{\o{\mathbbm{n}}{}^\alpha_A \cdot \o{\mathbbm{n}}{}^\beta_B}_{\o{\mathbbm{c}}{}^{\alpha \beta}_{AB}}
	+ \gamma^\alpha_A \gamma^\beta_A \underbrace{\o{\mathbbm{n}}{}^\alpha_A \cdot \o{\mathbbm{n}}{}^\beta_A}_{\o{\mathbbm{c}}{}^{\alpha \beta}_{AA}}\biggr] \, , 
\end{align*}
where $\vert \o{\b{g}} \vert$ is used to parametrize the grain-boundary free energy (cf.\ \eqn{G_norm_gurtin}). 
The intra- and inter-grain interaction moduli in the \citet{Beers2013} model are denoted by $\o{\mathbbm{c}}{}^{\alpha \beta}_{IJ}$.

A purported difference between the models of \citet{Gurtin2008} and \citet{Beers2013} is in the characterisation of the inter-grain interaction moduli, $\o{\mathbb{C}}{}^{\alpha \beta}_{AB}$ and $\o{\mathbbm{c}}{}^{\alpha \beta}_{AB}$.  
The relationship between  $\o{\b{G}}$ and  $\o{\b{g}}$ can be seen directly by assuming a planar problem in which $\b{l}^\alpha =-\b{e}_3$ and the slip plane and grain-boundary normal are in the $\b{e}_1$-$\b{e}_2$ plane. 
Screw dislocations cannot be accommodated in such a model as $\b{l}^\alpha_J \cdot \o{\b{n}} = 0$. 
Contracting the grain-boundary Burgers tensor \citep[using Eq.\ (9.8) in][]{Gurtin2008} with $-\b{e}_3$ gives  the net defect vector $\o{\b{g}}$:
\begin{align*}
-\o{\b{G}} \b{e}_3 &= 
\sum_\alpha \lrb{
	\gamma^\alpha_B [\b{s}^\alpha_B \cdot \o{\b{n}}][\b{s}^\alpha_B \otimes \b{l}^\alpha_B] 
	-\gamma^\alpha_B [\b{l}^\alpha_B \cdot \o{\b{n}}][\b{s}^\alpha_B \otimes \b{s}^\alpha_B] 
	-\gamma^\alpha_A [\b{s}^\alpha_A \cdot \o{\b{n}}][\b{s}^\alpha_A \otimes \b{l}^\alpha_A] 
	+\gamma^\alpha_A [\b{l}^\alpha_A \cdot \o{\b{n}}][\b{s}^\alpha_A \otimes \b{s}^\alpha_A] 
} \b{e}_3 \\
&= \sum_\alpha \lrb{
	\gamma^\alpha_B [\b{s}^\alpha_B \cdot \o{\b{n}}]\b{s}^\alpha_B 
	-\gamma^\alpha_A [\b{s}^\alpha_A \cdot \o{\b{n}}]\b{s}^\alpha_A
	} \\
	&=\o{\b{g}} \, .
\end{align*}
Hence the general theory of  \citet{Gurtin2008} reduces to that of \citet{Beers2013} under  planar restrictions.
Furthermore,
\begin{align*}
\o{\mathbb{C}}{}^{\alpha \beta}_{AB} \equiv \o{\mathbbm{c}}{}^{\alpha \beta}_{AB} = [\b{s}^\alpha_A \cdot \b{s}^\beta_B][\b{s}^\alpha_A\cdot\o{\b{n}}][\b{s}^\beta_B\cdot\o{\b{n}}] && \text{(planar problems)} \, .
\end{align*}

\begin{figure}[!ht]
 \centering
 \includegraphics[width = \textwidth]{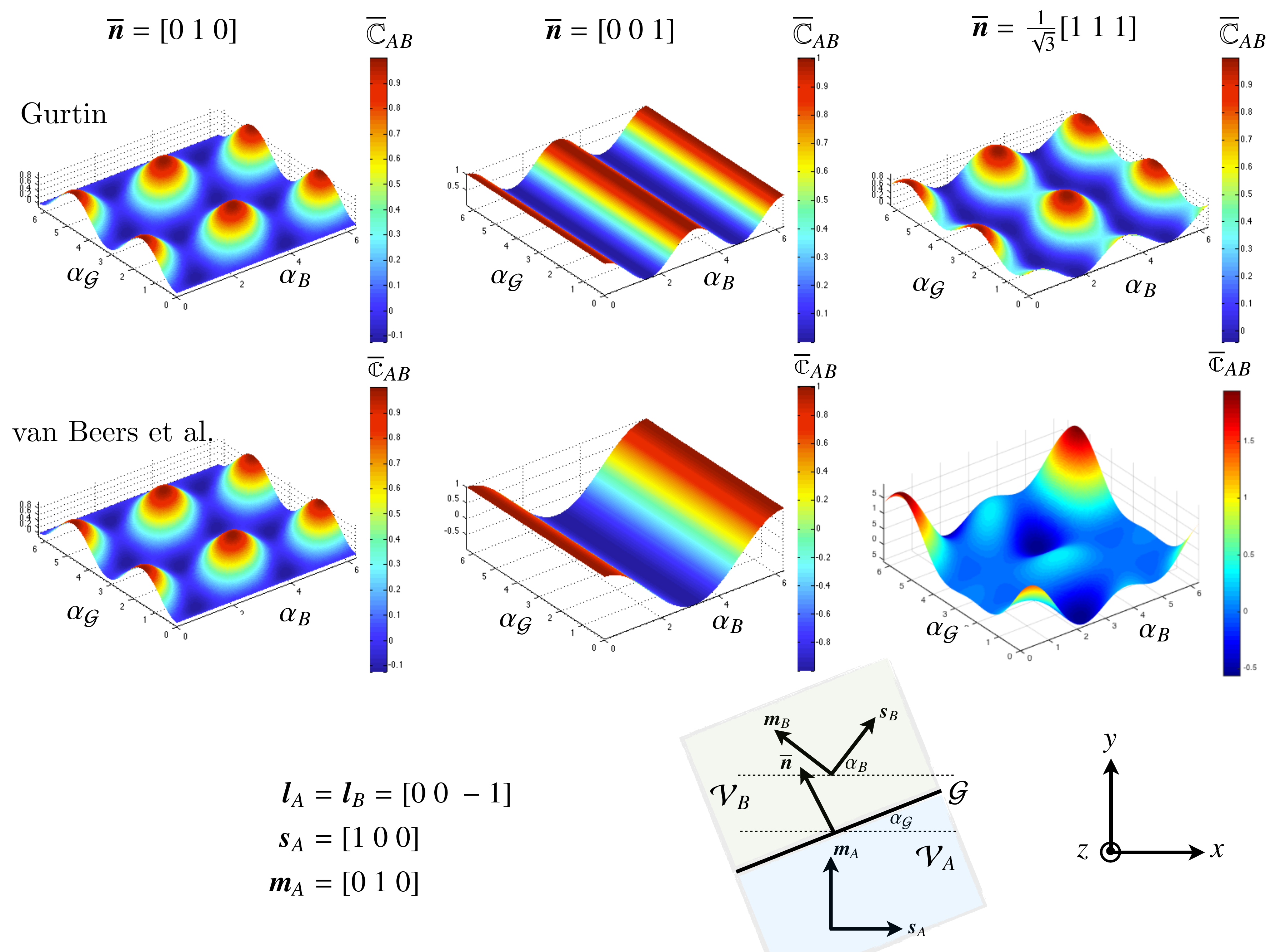}
 \caption{Comparison of the inter-grain interaction moduli in the models of \citet{Gurtin2008} and \citet{Beers2013} for a range of $\alpha_\mathcal{G}$ and  $\alpha_B$ [\si{\radian}]. 
 For the choice  $\o{\b{n}} = [0~1~0]$ the problem is planar.} 
 \label{gb_interaction}
\end{figure}

Consider the example of single slip in the bi-crystal shown in \fig{gb_interaction}. 
The slip system in grain A is fixed.
The slip system grain B is initially the same as A and is then rotated by an angle $\alpha_B$ around the $\b{e}_3$-axis. 
The grain-boundary normal is initially at the value shown in \fig{gb_interaction} and then rotated by an angle $\alpha_\mathcal{G}$ around the $\b{e}_3$-axis. 
Consider the planar case where the slip directions, slip plane normal, and the grain-boundary normal are constrained to lie in the $\b{e}_1$-$\b{e}_2$ plane.
As expected, the two theories produce identical inter-grain interaction moduli. 

The geometric arguments underpinning the vectorial quantity $\o{\b{g}}$ are insufficient when the problem becomes three dimensional, and the two theories differ.
The grain-boundary Burgers tensor $\o{\b{G}}$, however, is a fundamental measure that is well defined in three dimensions.  
It is clear from \fig{gb_interaction} that, as suggested by \citet{Gurtin2008}, the inter-grain interaction moduli vary continuously between two extremes $-1 \leq \o{\mathbb{C}}{}_{AB} \leq 1$ and correspond to physically intuitive notions of non-interactive and maximally interactive slip systems. 

The computed intra-grain interaction moduli for the planar problem are identical.
The intra-grain interaction moduli differ for non-planar problems. 


\section{Kinetics}\label{sec_kinetics}

The symmetric stress tensor in the bulk is denoted by $\b{T}$. 
Scalar and vector microscopic forces, denoted by $\pi^\alpha$ and $\b{\xi}^\alpha$ respectively, are postulated as conjugates to the slip rates and their spatial gradients \citep{Gurtin2000a, Gurtin2002}.
The vectorial microscopic stresses exist only in the bulk. 
The resolved shear stress  on slip system $\alpha$ is denoted by $\tau^\alpha := \b{T} : \mathbb{S}^\alpha$.
The power conjugate pairings are as follows:
\begin{align*}
\b{T} \leftrightarrow \dot{\b{E}}{}^\text{e} && \text{in } \mathcal{V}  && \text{(macroscopic stress),} \\
	\pi^\alpha 	\leftrightarrow \dot{\gamma}{}^\alpha && \text{in } \mathcal{V}  && \text{(scalar microscopic force),} \\ 	
	\b{\xi}^\alpha \leftrightarrow \grad \dot{\gamma}^\alpha && \text{in } \mathcal{V}	&& \text{(vector microscopic force)}  \, , \\ 
	\o{\pi}^\alpha_A \leftrightarrow \dot{\gamma}^\alpha_A \, ,
	\o{\pi}^\alpha_B \leftrightarrow \dot{\gamma}^\alpha_B  &&\text{on } \mathcal{G}_A, \mathcal{G}_B	&& \text{(scalar microscopic forces).}
\end{align*}
The scalar microforces on either side of the grain boundary are denoted by $\o{\pi}^\alpha_J$.

\subsection{Bulk}

The free energy in the bulk $\Psi$ is decoupled into an elastic (macroscopic) and a defect (microscopic) part, denoted by $\Psi^\text{e}$ and $\Psi^\text{d}$ respectively, given by\footnote{For further examples of defect energies and the inclusion of classical hardening in the free energy see \citet{Reddy2011}.}
\begin{align*}
	\Psi = \underbrace{\sfrac{1}{2}\b{E}^\text{e} : \mathcal{C}  \b{E}^\text{e}}_{\Psi^\text{e}} + \Psi^\text{d}(\u{\rho}) \, . 
\end{align*}

The defect energy $\Psi^\text{d}$ is chosen to be the following function of the dislocation densities $\u{\rho}$:
\begin{align*}
\Psi^\text{d} &= \sum_\alpha \frac{1}{2} \lrb{ C_\vdash {\rho_\vdash^{\alpha}}^2 + C_\odot {\rho_\odot^\alpha}^2} \, ,
\end{align*}
where the material parameters are given by \citep[see e.g.][]{Erturk2009, Evers2004}
\begin{align*}
C_\vdash = \dfrac{b^2 \mu R^2}{8[1-\nu]} 
&& \text{and} &&
C_\odot = \dfrac{b^2 \mu R^2}{4} \, ,
\end{align*}
and $R$ is the back stress cut-off radius and $\nu$ is the Poisson's ratio.
The dependence of $\Psi^\text{d}$  on the gradient of the slip is via relation (\ref{edge_screw}).

The elastic free energy  is chosen to be quadratic, for the sake of simplicity. 
The fourth-order isotropic elasticity tensor in the bulk is given by 
\begin{align*}
\mathcal{C}_{ijkl} =  \lambda \delta_{ij} \delta_{kl} 
	+ \mu \lrb{ \delta_{ik} \delta_{jl} + \delta_{il} \delta_{jk} } \, , 
\end{align*}
where $\lambda$ and $\mu$ are the Lam\'{e} constants.

The standard Coleman--Noll procedure gives the stress as the conjugate kinetic quantity to the elastic strain:
\begin{align}
\b{T} = \dfracp{ \Psi^\text{e}}{ \b{E}^\text{e}} = \mathcal{C} \b{E}^\text{e} \, . \label{elastic_constitutive}
\end{align}

The scalar energetic defect forces associated with edge and screw dislocations are respectively defined by
\begin{align*}
	f^\alpha_\vdash  = \dfracp{\Psi^\text{d}(\u{\rho})}{\rho^\alpha_\vdash} 
		&& \text{and} && 
	f^\alpha_\odot  = \dfracp{\Psi^\text{d}(\u{\rho})}{\rho^\alpha_\odot} \, .
\end{align*}
The energetic vectorial microstress $\b{\xi}_{\text{en}}^\alpha$ is defined by
\begin{align}
\b{\xi}_{\text{en}}^\alpha  
	= -f^\alpha_\vdash \b{s}^\alpha  + f^\alpha_\odot \b{l}^\alpha \, , \label{vec_microforce_constitutive}
\end{align}
and the reduced dissipation inequality in the bulk follows as
\begin{align*}
\mathcal{D}^\text{red} &:=  \sum_\alpha \biggl[ \underbrace{\lrb{ \b{\xi}^\alpha -\b{\xi}_\text{en}^\alpha}}_{\b{\xi}_\text{dis}^\alpha} \cdot \grad \dot{\gamma}{}^\alpha 
	 +  \pi^\alpha \dot{\gamma}{}^\alpha  \biggr] \geq 0 \, ,
\end{align*}
where $\b{\xi}_\text{dis}^\alpha$ is the dissipative vectorial microstress.
This general framework is restricted (to elucidate the role of the grain boundary) to purely energetic vectorial microstresses  and purely dissipative scalar microforces:
\begin{align}
	\b{\xi}_\text{dis}^\alpha \equiv \b{0}	&& \implies && \b{\xi}^\alpha \equiv \b{\xi}_\text{en}^\alpha \notag \\
		\pi_\text{en}^\alpha \equiv 0	&& \implies && \pi^\alpha \equiv \pi_\text{dis}^\alpha \notag \\
\implies &&	\mathcal{D}^\text{red} \equiv  \sum_\alpha \pi^\alpha \dot{\gamma}{}^\alpha \geq 0	 \, . \label{red_diss_volume}
\end{align}

\subsection{Grain boundary}

Following the general theory of \citet{Gurtin2008}, we  consider both energetic and dissipative structures in the grain boundary. 
The influence of these two structures was numerically investigated by \citet{Beers2013}. 

The free energy of the grain boundary is parametrized in terms of the grain-boundary Burgers tensor, that is $\o{\Psi} = \o{\Psi}(\o{\b{G}})$.
The grain-boundary \emph{energetic} stress is defined by
\begin{align}
\o{\b{M}} := \dfracp{\o{\Psi}(\o{\b{G}})}{\o{\b{G}}} \, , \notag
\intertext{thus}
\dot{\o{\Psi}} = \o{\b{M}} : \dot{\o{\b{G}}} \, . \notag 
\end{align}
Following \citet{Gurtin2008}, the grain-boundary free energy is assumed to be given by the  quadratic function
\begin{align*}
\o{\Psi}(\o{\b{G}}) = \sfrac{1}{2} \o{\lambda} \vert \o{\b{G}} \vert^2 \, ,
\intertext{which yields}
\o{\b{M}} = \o{\lambda} \, \o{\b{G}} \, ,
\end{align*}
where $\o{\lambda} > 0$ is a constant constitutive parameter. 
The rate of change of the grain-boundary free energy can be expressed in terms of the slip rates on either side of the grain boundary, using \eqn{G_bar_dot}, as follows:
\begin{align*}
\dot{\o{\Psi}} = \o{\b{M}} : \dot{\o{\b{G}}}	&=  \o{\lambda} \, \o{\b{G}} : \dot{\o{\b{G}}} \\
	&=   \sum_\alpha \lrb{  [\o{\lambda}\,  {\o{\mathbb{N}}{}^\alpha_B}:\o{\b{G}}]\dot{\gamma}{}^\alpha_B  
		- [ \o{\lambda}\,  {\o{\mathbb{N}}{}^\alpha_A}:\o{\b{G}}]\dot{\gamma}{}^\alpha_A  } \\
		&=  \sum_\alpha  \lrb{\o{\pi}{}^\alpha_{B,\text{en}} \dot{\gamma}{}^\alpha_B
		+ \o{\pi}{}^\alpha_{A,\text{en}} \dot{\gamma}{}^\alpha_A} \, ,
\end{align*}
 where the \emph{energetic internal microforces} are defined by 
$\o{\pi}{}^\alpha_{B,\text{en}} :=  \o{\lambda}\,  {\o{\mathbb{N}}{}^\alpha_B}:\o{\b{G}}$ and 
$\o{\pi}{}^\alpha_{A,\text{en}} :=  -\o{\lambda}\,  {\o{\mathbb{N}}{}^\alpha_A}:\o{\b{G}}$.

Following \citet{Gurtin2005, Gurtin2008}, the reduced dissipation on the grain boundary is expressed in terms of the rate of change of the grain-boundary Burgers tensor  $\dot{\o{\b{G}}}$ and its conjugate kinetic quantity $\o{\b{K}}$:
\begin{align}
\o{\mathcal{D}}{}^\text{red} =  \sum_\alpha \biggl[
 	\underbrace{[\o{\pi}{}^\alpha_A - \o{\pi}{}^\alpha_{A,\text{en}}]}_{\o{\pi}{}^\alpha_{A,\text{dis}}} \dot{\gamma}{}^\alpha_A 
 + 	\underbrace{[\o{\pi}{}^\alpha_B  - \o{\pi}{}^\alpha_{B,\text{en}}]}_{\o{\pi}{}^\alpha_{B,\text{dis}}}
 \dot{\gamma}{}^\alpha_B
\biggr] 
  = 
 \o{\b{K}} : \dot{\o{\b{G}}} \geq 0 	 \, , \label{red_diss_GB}
\end{align} 
where the dissipative internal microscopic forces on the grain boundary $\o{\pi}_{A,\text{dis}}$ and $\o{\pi}_{B,\text{dis}}$ are defined by the relations
\begin{align}
\o{\pi}{}^\alpha_{A,\text{dis}} = - \o{\mathbb{N}}{}^\alpha_A : \o{\b{K}}
&& \text{and} &&
\o{\pi}{}^\alpha_{B,\text{dis}} =   \o{\mathbb{N}}{}^\alpha_B : \o{\b{K}} \, . \label{pi_K_relation}
\end{align}

\section{Governing equations}\label{sec_governing_equations}

The equations governing the response of the bulk and the grain boundary are now summarised.

\subsection{Bulk}

A balance of macroscopic and microscopic forces in the bulk yields 
\begin{align}
 \div \b{T}  = \b{0} \quad \text{in } \mathcal{V} 
 && \text{and} && 
 \b{t}^\star(\b{n}) = \b{T}  \b{n} \quad \text{on } \partial \mathcal{V}_N \, ,  
 \label{macro_force_balance_strong}\\
 \div \b{\xi}^\alpha + \tau^\alpha - \pi^\alpha = 0 \quad \text{in } \mathcal{V} 
 && \text{and} &&
 \Xi^{\alpha\star}(\b{n}) = \b{\xi}^\alpha\cdot \b{n} \quad \text{on } \partial \mathcal{V}_F \, .
 \label{micro_force_balance_strong}
\end{align}
\eqn{macro_force_balance_strong} is the standard equilibrium equation in the absence of body and inertial  forces, and $\b{t}^\star$ is the prescribed Cauchy traction on the Neumann part of the boundary $\partial \mathcal{V}_N$.
Dirichlet boundary conditions on the displacement  $\b{u}$ are prescribed on $\partial \mathcal{V}_D$, where $\partial \mathcal{V} = \partial \mathcal{V}_N \cup \partial \mathcal{V}_D$ and $\partial \mathcal{V}_N \cap \partial \mathcal{V}_D = \emptyset$.
Furthermore, the boundary $\partial \mathcal{V}$ is  subdivided into complementary parts $\partial \mathcal{V}_F$ and $\partial \mathcal{V}_H$. 
The standard boundary condition on the micro-free part of the boundary  $\partial \mathcal{V}_F$ is that the scalar microscopic traction $\Xi^{\alpha\star} = 0$ while on the micro-hard part of the boundary $\partial \mathcal{V}_H$ homogeneous conditions on the slip are prescribed.  
For additional details on the microscopic boundary conditions see \citet{Gurtin2005}. 
The macroscopic and microscopic force balances are coupled via the dependence of the resolved shear stress $\tau^\alpha$ on the macroscopic stress $\b{T}$.

\subsection{Grain boundary}

The balance of macroscopic and microscopic forces across the grain boundary yields 
\begin{align}
\jmp{\b{T}}  \o{\b{n}} = \b{0} \quad \text{on } \mathcal{G} \label{macroforce_gb}\\
 -\o{\pi}^\alpha_A - \b{\xi}^\alpha_A \cdot \o{\b{n}} = 0  \quad \text{on } \mathcal{G}_A
	&& \text{and} &&
 - \o{\pi}^\alpha_B + \b{\xi}^\alpha_B \cdot \o{\b{n}} = 0  \quad \text{on } \mathcal{G}_B \label{microforce_gb}\, .
\end{align}
\eqn{macroforce_gb} is the standard traction continuity condition for an interface.
The microforce balance (\ref{microforce_gb}) on the grain boundary states that the internal microforce on either side of the grain boundary acts in response to the flux of the vectorial microforce from the grain. 
The gradient-plasticity formulation adopted in the bulk allows meaningful balance equations for the grain boundary to be constructed in a consistent manner. 
A micro-hard or micro-free boundary condition could be applied on the grain boundary. 
However all information about the geometry of neighbouring crystal structures and grain-boundary orientation would be lost. 

\section{Plastic flow relations}\label{sec_flow_relations}

In order to complete the theory, the reduced dissipation inequalities in the various parts of the body (\ref{red_diss_volume}, \ref{red_diss_GB})  need to be satisfied in a thermodynamically consistent manner.
The plastic flow relations are then obtained by postulating forms for the dissipative internal variables.

\subsection{Bulk}

The yield function $f(\pi^\alpha)$ defines the  region of admissible microscopic forces on the $\alpha^\text{th}$ slip surface.
The yield function and the flow law for the plastic slip are defined by
\begin{align}
f(\pi^\alpha) &= \vert \pi^\alpha \vert - S \leq 0 \, , \label{yield_surface_bulk} \\
\dot{\gamma}{}^\alpha &= \lambda^\alpha \dfracp{f(\pi^\alpha)}{\pi^\alpha} = \lambda^\alpha \text{sgn} \pi^\alpha \, , \label{rate_indep_flow_bulk}
\end{align}
where $S >0 $ is the constant slip resistance and $\lambda^\alpha \geq 0$ is a scalar multiplier, together with the Kuhn--Tucker complementarity conditions
\begin{align}
f(\pi^\alpha) \leq 0 \, ,
&&
\lambda^\alpha \geq 0 \, , 
&& 
\lambda^\alpha f(\pi^\alpha) = 0 \, . \label{KTT}
\end{align}
Under conditions of plastic flow $f(\pi^\alpha) \equiv 0$ and the flow rule (\ref{rate_indep_flow_bulk}) can be inverted to obtain
\begin{align}
	\pi^\alpha &= 
	 S \text{sgn} \dot{\gamma}{}^\alpha 
	= \dfracp{D_\text{eff}^\alpha}{\dot{\gamma}{}^\alpha}
 \, , \label{pi_rate_independent}
\end{align} 
where the effective dissipation function $D_\text{eff}^\alpha$ is given by $S \vert \dot{\gamma}{}^\alpha \vert$. 
More generally, the elastic region, the flow rule and the complementarity conditions  can be expressed in the alternative, equivalent form 
\begin{align*}
	D_\text{eff}^\alpha(\t{\gamma}^\alpha) \geq D_\text{eff}^\alpha(\dot{\gamma}{}^\alpha)
	+ \pi^\alpha [ \t{\gamma}^\alpha- \dot{\gamma}{}^\alpha ] \, .
\end{align*}
This formulation, which has received a detailed treatment in \citet{Han2013}, is less popular as a basis for computational treatments of the classical problem than (\ref{yield_surface_bulk})--(\ref{KTT}), which uses the flow law in its traditional form of the normality law with the Kuhn--Tucker conditions. 
Nevertheless this formulation is particularly well suited to problems such as  gradient plasticity in which the plastic slip is a primary unknown.

The rate-independent theory presents various numerical challenges due to the indeterminacy of plastic slip.
To circumvent these problems, a regularized effective dissipation function $D^\alpha_\text{vis}$ is proposed of the form
\begin{align*}
D^\alpha_\text{vis} = \dfrac{S}{m+1} \lrb{ \dfrac{\vert \dot{\gamma}{}^\alpha \vert}{\dot{\gamma}{}_0}}^{m+1}\dot{\gamma}{}_0 \, ,
\end{align*}
where $\dot{\gamma}{}_0$ is the reference value for the slip rate and  $m>0$ is the rate sensitivity. 
The scalar microforce follows as (cf.\ \eqn{pi_rate_independent}):
\begin{align}
\pi^\alpha = \dfracp{D_\text{vis}^\alpha}{\dot{\gamma}{}^\alpha} 
 = S \lrb{ \dfrac{\vert \dot{\gamma}{}^\alpha \vert}{\dot{\gamma}_0}}^m \text{sgn} \dot{\gamma}{}^\alpha \, . \label{plastic_flow_bulk}
\end{align}

\subsection{Grain boundary}\label{sec:grain_boundary_flow_relations}

The grain boundary impedes the flow of dislocations from the adjoining grains. 
The geometrical structure of the grain boundary quantifies the amount of impedance. 
Should a measure of the microforce on the  grain boundary reach a sufficient level, the grain boundary can yield and allow for transmission of dislocations. 

\citet{Gurtin2008} proposes two forms for the flow relation on the grain boundary, both of which satisfy the reduced dissipation inequality on the grain boundary (\ref{red_diss_GB}). 
The first is a flow relation explicitly parametrized by $\o{\b{G}}$ (here referred to as the \citeauthor{Gurtin2008}~I model). 
The flow relation therefore accounts for the interaction of the slip systems in adjacent grains and the grain boundary orientation. 
The second  relation is parametrized in terms of the slip on either side of the grain boundary (here referred to as the \citeauthor{Gurtin2008}~II model). 

The \citeauthor{Gurtin2008}~I model allows for the interaction of slips from either side of the grain boundary via the definition (\ref{G_bar_dot}) of $\dot{\o{\b{G}}}$. 
It is thus possible to have a $\dot{\o{\b{G}}} \equiv \b{0}$, and hence zero dissipation, for non-zero values of the slip rate $\dot{\u{\gamma}}{}_I$. 
The \citeauthor{Gurtin2008}~II model excludes the possibility of non-zero slip resulting in zero dissipation. 
A modified definition for $\o{\b{G}}$ is given in \sect{sec:mod_G} that, when used to parametrize the flow rule, ensures that $\dot{\o{\b{G}}} \equiv \b{0}$ if and only if $\dot{\u{\gamma}}{}_I \equiv 0$. 
The response of the three models are explored via a series of numerical examples in \sect{sec_numerical_examples}.

\subsubsection{Plastic flow relation in terms of $\o{\b{G}}$ (Gurtin I model)}
 
A regularized dissipation function for the grain boundary is chosen as 
\begin{align}
\o{D}_\text{vis} = 	\dfrac{\o{S}}{\o{m}+1} \lrb{ \dfrac{ \vert \dot{\o{\b{G}}} \vert}{\dot{\o{{G}}}_0}}^{\o{m}+1} \dot{\o{{G}}}_0 \, , \label{D_vis_G}
\end{align}
where $\o{S} > 0 $ is the slip resistance, $\dot{\o{G}}{}_0$ is the reference value for the rate of the grain-boundary Burgers tensor and $\o{m} > 0$ is the rate sensitivity parameter. 
The flow relation for $\o{\b{K}}$ follows as 
\begin{align}
\o{\b{K}} = \dfracp{\o{D}_\text{vis}}{\dot{\o{\b{G}}}} 
=
\overline{S}
	\lrb{\dfrac{\vert \dot{\o{\b{G}}} \vert}{\dot{\o{{G}}}_0}}^{\o{m}}
	\dfrac{\dot{\o{\b{G}}}}{ \vert \dot{\o{\b{G}}} \vert} \, . \label{plastic_flow_grain_boundary}
\end{align}
It follows from \eqn{pi_K_relation} that 
\begin{align*}
\o{\pi}{}^\alpha_{A,\text{dis}} = - \overline{S}
	\lrb{\dfrac{\vert \dot{\o{\b{G}}} \vert}{\dot{\o{{G}}}_0}}^{\o{m}}
	\dfrac{{\o{\mathbb{N}}{}^\alpha_A} : \dot{\o{\b{G}}}}{ \vert \dot{\o{\b{G}}} \vert}
	&& \text{and} && 
\o{\pi}{}^\alpha_{B,\text{dis}} =  \overline{S}
	\lrb{\dfrac{\vert \dot{\o{\b{G}}} \vert}{\dot{\o{{G}}}_0}}^{\o{m}}
	\dfrac{{\o{\mathbb{N}}{}^\alpha_B} : \dot{\o{\b{G}}}}{ \vert \dot{\o{\b{G}}} \vert} \, .
\end{align*}
The above relations can be expressed in terms of the slip rates using \eqn{G_bar_dot}.

\subsubsection{Plastic flow relation in terms of $\dot{{\gamma}}$ (Gurtin II model)}

A regularized dissipation function for side $I$ of the grain boundary is chosen as 
\begin{align*}
\o{D}^\alpha_{\text{vis},I} = 	\dfrac{\o{S}}{\o{m}+1} \lrb{ \dfrac{ \vert \dot{{\gamma}}_I^\alpha \vert}{\dot{\o{\gamma}}_0}}^{\o{m}+1} \dot{\o{{\gamma}}}_0 \, ,
\end{align*}
where $\dot{\o{\gamma}}{}_0$ is the reference value for the rate of the grain-boundary slip, and the remaining parameters are defined as in \eqn{D_vis_G}. 
The regularized dissipation function for the grain boundary follows as
\begin{align*}
\o{D}^\gamma_\text{vis} = \sum_\alpha \lrb{\o{D}^\alpha_{\text{vis},A} + \o{D}^\alpha_{\text{vis},B}} \, .
\end{align*}
The dissipative internal microscopic forces on either side of the grain boundary (see \eqn{microforce_gb}) are thus obtained as
\begin{align*}
\o{\pi}{}^\alpha_{I,\text{dis}} 
=  \dfracp{\o{D}^\gamma_{\text{vis}}}{\dot{{\gamma}}{}^\alpha_I}
= \overline{S}
	\lrb{\dfrac{\vert \dot{{\gamma}}^\alpha_I \vert}{\dot{\o{{\gamma}}}_0}}^{\o{m}}
	\dfrac{\dot{{\gamma}}^\alpha_I}{ \vert \dot{{\gamma}}^\alpha_I \vert} \, .
\end{align*}

\subsection{A modified form for the flow relation on the grain boundary}\label{sec:mod_G}

Consider the \citeauthor{Gurtin2008b}~I model.
It is possible to construct situations such that 
\begin{equation}
\o{\b{G}} = \sum_\alpha \o{\b{G}}{}^\alpha =  \sum_\alpha \lrb{ \gamma^\alpha_B \o{\mathbb{N}}{}^\alpha_B - \gamma^\alpha_A \o{\mathbb{N}}{}^\alpha_A} \equiv \b{0}\,
\label{G=0}
\end{equation}
for non-zero values of $\gamma^\alpha$ on both sides of the grain boundary (see \ref{app_gurtin_I} for an illustrative example). 
Such situations are illustrated in \sect{sec:multi_slip} for a three-dimensional example with a face-centered-cubic crystal structure. 
This situation renders the expression (\ref{red_diss_GB}) for the corresponding dissipation meaningless.
It should be emphasised that this is not a deficiency of the \citeauthor{Gurtin2008b}~I model. 
It could be argued that the recombination of slip in the grain boundary accounts for physical phenomena such as annihilation.
As demonstrated in \sect{sec_numerical_examples}, the model is not, however, capable of reproducing the behaviour of both the micro-hard and micro-free boundary conditions. 
With this as motivation we revisit the definition of the grain-boundary Burgers tensor with a view to constructing an alternative measure of ``defect" density across the boundary.

The first observation is that the Burgers tensor in the bulk $\b{G}$ is obtained from the plastic deformation tensor $\b{H}^\text{p}$, which by \eqn{H_p} involves a sum over all slip systems: the derivation of $\b{G}$ involves taking a circuit of $\b{H}^\text{p}$ and then using Stokes' Theorem. 
The derivation of the grain-boundary equivalent involves taking an  infinitesimal circuit that includes a section of the grain boundary. 
In contrast to the situation in the bulk, the slip systems on either side of the boundary bear no relation to each other. 
In particular, there is no relationship between the numbering adopted on either side. 
So instead of considering the jump in $\b{H}^\text{p}$, that is,
\begin{align*}
\o{\b{H}}\prs = \jmp{ \b{H}^\text{p} } &= 
\sum_\alpha \lrb{\gamma^\alpha_B \mathbb{S}{}^\alpha_B  - \gamma^\alpha_A \mathbb{S}{}^\alpha_A} \, ,
\end{align*}
we consider the differences between the individual components that make up $\b{H}^\text{p}$. 

By analogy with the definition of $\o{\b{G}}{}^\alpha$ consider the difference 
\begin{align*}
\h{\b{G}}{}^{\alpha\beta}_R 
& := \sqrt{R^{\alpha\beta}}\lrb{\gamma^\beta_B\o{\mathbb{N}}{}^\beta_B - \gamma^\alpha_A\o{\mathbb{N}}{}^\alpha_A} \,.
\end{align*}
Here $R^{\alpha\beta}$ is a flag whose components take the values 1 or 0, in this way allowing a selection of particular combinations of slip systems for the two adjacent grains. 
The square root is introduced in order to arrive at a convenient formulation for the magnitude later.
Then
\begin{align}
\bigl|\h{\b{G}}{}^{\alpha\beta}_R \bigr|^2 & =  
	R^{\alpha\beta}
\bigl| \gamma^\beta_B \o{\mathbb{N}}{}^\beta_B - \gamma^\alpha_A\o{\mathbb{N}}{}^\alpha_A \bigr|^2  \,.
\label{absGhatab}
\end{align}
We note also from \eqn{G=0} that 
\begin{equation*}
\o{\b{G}} = \sum_\alpha \h{\b{G}}{}^{\alpha\alpha}_\delta \, ,
\end{equation*}
in which the subscript $\delta$ denotes the choice $R^{\alpha\beta} = \delta_{\alpha\beta}$.

Now define a modified magnitude $\Gamma_R$ of $\o{\b{G}}$ by
\begin{align*}
\Gamma^2_R & = \sum_{\alpha,\beta} 
\bigl|\h{\b{G}}{}^{\alpha\beta}_R \bigr|^2 \, .
\end{align*}
This may be compared with the conventional magnitude which is given by
\begin{align*}
|\o{\b{G}}|^2 & = \biggl| \sum_\alpha \h{\b{G}}{}^{\alpha\alpha}_\delta \biggr|^2 \notag \\
& = \biggl| \sum_\alpha \lrb{ \gamma_B^\alpha\o{\mathbb{N}}{}_B^\alpha - \gamma_A^\alpha\o{\mathbb{N}}{}_A^\alpha} \biggr|^2\,.
\end{align*}
Thus we see that the conventional magnitude is the absolute value of a sum while the modification is the sum of weighted absolute values.

We have
\begin{align*}
|\o{\b{G}}|^2 & = \biggl| \sum_\alpha \h{\b{G}}{}^{\alpha\alpha}_\delta \biggr|^2 
\leq 
 \sum_\alpha   \bigl| \h{\b{G}}{}^{\alpha\alpha}_\delta  \bigr|^2
=
  \sum_\alpha \bigl| \gamma_B^\alpha\o{\mathbb{N}}{}_B^\alpha - \gamma_A^\alpha\o{\mathbb{N}}{}_A^\alpha] \bigr|^2 
  = \sum_{\alpha,\beta} \bigl|\h{\b{G}}{}^{\alpha\beta}_\delta \bigr|^2 \, .
\end{align*}
Note also that for the special case in which $R^{\alpha\beta} = \delta_{\alpha\beta}$, with $\delta_{\alpha\beta}$ being the Kronecker delta,
\begin{align*}
\Gamma^2_\delta & = \sum_\alpha \big|\h{\b{G}}{}^{\alpha\alpha}_\delta\big|^2 \,.
\end{align*}
Thus
\begin{align*}
|\o{\b{G}}|^2 \leq \Gamma^2_\delta \,.
\end{align*}

\subsection{The dissipation function}\label{sec:dissfnG}
The modified definition of the magnitude of $\o{\b{G}}$ alters the form of  the dissipation inequality and flow rule on the grain boundary. 
In what follows we drop the subscript $R$ on $\h{\b{G}}{}^{\alpha\beta}_R$ for convenience. 
Furthermore we consider only dissipative microforces on the grain boundary.

The reduced dissipation inequality on the grain boundary (\ref{red_diss_GB}) is restated as
\begin{align*}
\o{\mathcal{D}}{}^\text{red} =  
\sum_\alpha  	\o{\pi}{}^\alpha_A \dot{\gamma}{}^\alpha_A 
 + 	
\sum_\beta \o{\pi}{}^\beta_B \dot{\gamma}{}^\beta_B
  = 
 \o{\b{K}} : \dot{\o{\b{G}}} \geq 0 	 \, .
\end{align*} 
Following a similar procedure that lead to the definition of $\o{\b{K}}$, we write
\begin{align*}
\o{\mathcal{D}}{}^\text{red}  & =  
\sum_\alpha  	\o{\pi}{}^\alpha_A \dot{\gamma}{}^\alpha_A 
 + 	
\sum_\beta \o{\pi}{}^\beta_B \dot{\gamma}{}^\beta_B
  = 
 \sum_{\alpha , \beta} \h{\b{K}}{}^{\alpha \beta} : \dot{\h{\b{G}}}{}^{\alpha \beta} \geq 0 \, ,
\end{align*}
where $\h{\b{K}}{}^{\alpha \beta}$ has to be defined. 
Expand this expression to obtain
\begin{align*}
\o{\mathcal{D}}{}^\text{red}  & =  
 \sum_{\alpha , \beta} \h{\b{K}}{}^{\alpha \beta} : 
	\sqrt{R^{\alpha\beta}}
\lrb{ \dot{\gamma}^\beta_B \o{\mathbb{N}}{}^\beta_B - \dot{\gamma}^\alpha_A\o{\mathbb{N}}{}^\alpha_A }   \\
& =  
\sum_\beta \underbrace{\biggl[\sum_\alpha \sqrt{R^{\alpha\beta}} \h{\b{K}}{}^{\alpha \beta} : \o{\mathbb{N}}{}^\beta_B  \biggr]}_{\h{\pi}{}^\beta_B} \dot{\gamma}^\beta_B
- \sum_\alpha
\underbrace{\biggl[\sum_\beta \sqrt{R^{\alpha\beta}} \h{\b{K}}{}^{\alpha \beta} : \o{\mathbb{N}}{}^\alpha_A  \biggr]}_{-\h{\pi}{}^\alpha_A} \dot{\gamma}^\alpha_A \, ,
\end{align*}
where $\h{\pi}{}^\beta_B$ and $\h{\pi}{}^\alpha_A$ are the modified scalar microforces on adjacent sides of the grain boundary.

For the rate-independent case the modified yield function $\h{f}$ is defined by
\begin{equation*}
\h{f}(\h{\b{K}}{}^{\alpha\beta} ) = \biggl[ \sum_{\alpha,\beta}  \bigl| \h{\b{K}}^{\alpha\beta} \bigr|^2 \biggr]^{1/2} - S \leq 0\,.
\end{equation*}
Then the normality relation is
\begin{align*}
\dot{\h{\b{G}}}{}^{\alpha \beta} &= 
\h{\lambda} \dfracp{\h{f}(\h{\b{K}}{}^{\alpha\beta} )}{\h{\b{K}}{}^{\alpha\beta}}
= \h{\lambda} \dfrac{\h{\b{K}}{}^{\alpha\beta}}{ \bigl| \h{\b{K}}{}^{\alpha\beta} \bigr|} \, . 
\end{align*}

The corresponding dissipation potential is
\begin{align*}
\h{\mathcal{D}}(\dot{\h{\b{G}}}{}^{\alpha \beta}) &=
S\dot{\Gamma} 
=
S \biggl[ \sum_{\alpha,\beta}  \bigl| \dot{\h{\b{G}}}^{\alpha\beta} \bigr|^2 \biggr]^{1/2} \, ,
\end{align*}
so that when flow occurs
\begin{align*}
\h{\b{K}}{}^{\alpha\beta} &= 
\dfracp{\h{\mathcal{D}}(\dot{\h{\b{G}}}{}^{\alpha \beta})}{\dot{\h{\b{G}}}{}^{\alpha \beta}} 
= S \dfrac{\dot{\h{\b{G}}}{}^{\alpha \beta}}{\dot{\Gamma}} \, .
\end{align*}

\begin{pot}
The modified and original formulations are identical for the case of single slip with $R^{\alpha \beta}=1$.
The choice made here for $R^{\alpha \beta}$ in multi-slip problems is as follows.
For each slip system $\alpha$ in grain A, the most interactive slip system in grain B is identified.
The most interactive slip system in grain B is the slip system most likely to receive dislocations transmitted from slip system $\alpha$ in grain A, should plastic flow occur.
The process is repeated for grain B. 
The simple example in the following section provides further details.
Many other choices are possible; the choice made here is based on the concept of inter-grain interaction defined in the original theory of \citet{Gurtin2008} and ensures that the dissipation at the grain boundary is zero if and only if the slip on all adjacent slip systems is zero.  
The modified model  contains features of both the \citet{Gurtin2008}~I and II models. 
\end{pot}

\subsection{An illustrative example}\label{sec:example_modG}

\begin{figure}[!ht]
 \centering
 \includegraphics[width = 0.8\textwidth]{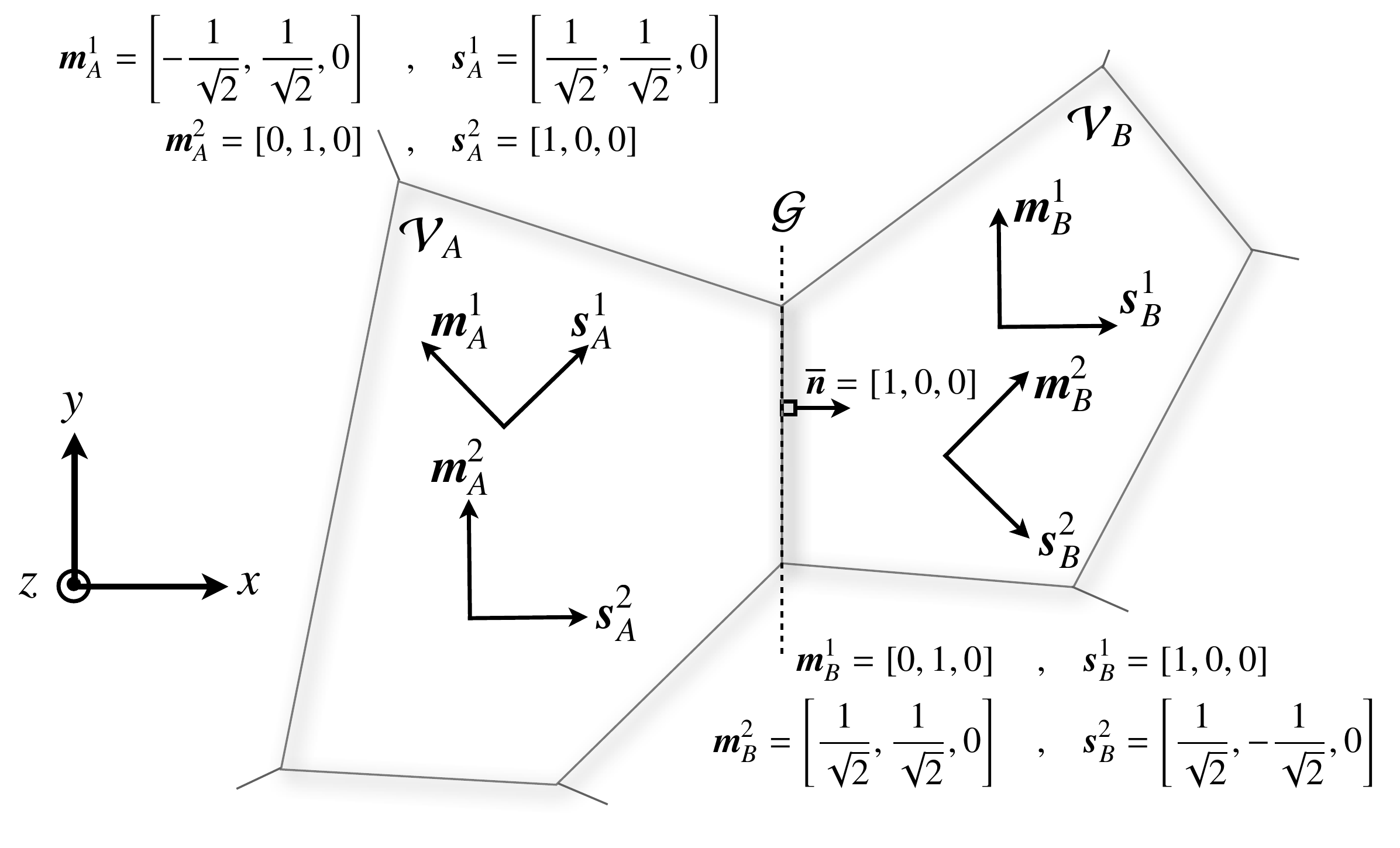}
 \caption{Illustrative example of two grains $\mathcal{V}_A$ and $\mathcal{V}_B$ separated by a grain boundary $\mathcal{G}$. }
 \label{mod_gamma}
\end{figure}

The procedure to select the most interactive slip system in adjacent grains, $\mathcal{V}_A$ and $\mathcal{V}_B$,  separated by a grain boundary $\mathcal{G}$ and hence compute $R^{\alpha \beta}$ is illustrated using the simple example problem shown in \fig{mod_gamma}. 
The matrix of inter-grain interaction moduli $\o{\mathbb{C}}_{AB}$ is  computed using \eqn{C_AB_Gurtin} as 
\begin{align*}
\o{\mathbb{C}}_{AB} = 
\begin{bmatrix}
0.5 & 0 \\
1 & 0.5
\end{bmatrix} \, 
&& \text{and} && 
\o{\mathbb{C}}_{BA} = 
\begin{bmatrix}
0.5 & 1 \\
0 & 0.5
\end{bmatrix} \, .
\end{align*} 
Recall that from the symmetry relation (\ref{interaction_sym}) $\o{\mathbb{C}}_{AB} = \o{\mathbb{C}}{}\trns_{BA}$.
The matrix $\o{\mathbb{C}}_{AB}$, with components $[\o{\mathbb{C}}_{AB}]^{\alpha \beta}$, contains the interaction coefficients for all slip systems $\alpha$ in $\mathcal{V}_A$ with all other slip systems $\beta$ in $\mathcal{V}_B$. 

Consider a slip system $\alpha$ in  $\mathcal{V}_A$. 
The slip system in  $\mathcal{V}_B$ that interacts the most with $\alpha$ in  $\mathcal{V}_A$ is given by the choice of $\beta$ that gives the maximum value of  $[\o{\mathbb{C}}_{AB}]^{\alpha\beta}$. 
For example, slip system 1 in $\mathcal{V}_B$ interacts the most with slip system 2 in $\mathcal{V}_A$. 
Consider now slip system $\alpha$ in  $\mathcal{V}_B$. 
Slip system 2 in $\mathcal{V}_B$ interacts the most with slip system 2 in $\mathcal{V}_A$. 
The corresponding interaction matrices $R_{IJ}$ have a value of 1 in the $\alpha \beta$ slot if slip system $\beta$ in $\mathcal{V}_J$ interacts the most with slip system $\alpha$ in $\mathcal{V}_I$, and 0 otherwise.
The interaction matrices for each grain are 
\begin{align*}
\mathcal{R}_{AB} = 
\begin{bmatrix}
1 & 0 \\
1 & 0
\end{bmatrix} 
&& \text{and} &&
\mathcal{R}_{BA} = 
\begin{bmatrix}
0 & 1 \\
0 & 1
\end{bmatrix} 
\,.
\end{align*}
The interaction matrix for the grain boundary $\mathcal{R}$ is defined by
\begin{align*}
\mathcal{R} &= \mathcal{R}_{AB} + \mathcal{R}\trns_{BA}  \, .
\end{align*}

The coefficients of the matrix $R$ in \eqn{absGhatab} are obtained from $\mathcal{R}$ as follows:
\begin{align*}
R^{\alpha\beta} &= \text{min}(1,\mathcal{R}_{\alpha\beta}) \, , \\
\implies R &= 
 \begin{bmatrix}
1 & 0 \\
1 & 1
\end{bmatrix}\, .
\end{align*}
Thus, for the example problem, $\Gamma^2_R$ is given by
\begin{align*}
\Gamma^2_R & = \sum_{\alpha,\beta} 
R^{\alpha\beta}
\bigl| \gamma^\beta_B \o{\mathbb{N}}{}^\beta_B - \gamma^\alpha_A\o{\mathbb{N}}{}^\alpha_A \bigr|^2  \\
&= 
1\bigl| \gamma^1_B \o{\mathbb{N}}{}^1_B - \gamma^1_A\o{\mathbb{N}}{}^1_A \bigr|^2 
+ 0\bigl| \gamma^2_B \o{\mathbb{N}}{}^2_B - \gamma^1_A\o{\mathbb{N}}{}^1_A \bigr|^2 \\
& \quad 
1\bigl| \gamma^1_B \o{\mathbb{N}}{}^1_B - \gamma^2_A\o{\mathbb{N}}{}^2_A \bigr|^2 
+ 1\bigl| \gamma^2_B \o{\mathbb{N}}{}^2_B - \gamma^2_A\o{\mathbb{N}}{}^2_A \bigr|^2 \, .
\end{align*}

\section{Extension of the variational problem to include grain boundaries}\label{sec_variational_problem}

Variational formulations of the \citet{Gurtin2002} model for strain-gradient, single-crystal plasticity are investigated in \citet{Reddy2011,Gurtin2014} \citep[see][for the polycrystalline case]{Reddy2011a}. 
The rate-dependent and independent cases are considered.
The variational formulation developed in \citet{Reddy2011} is extended here to include the grain boundary. 
The associated incremental minimization problem, discussed in \sect{minprob}, provides the point of departure for the numerical implementation in \sect{sec_fem}. 

\subsection{The variational problem}

The spaces of displacement $V$ and slips $Q$ are defined by
\begin{align*}
V &= \biggl\{  \b{u}~:~u_i, \, \dfracp{ u_i}{x_j} \in L^2(\mathcal{V}), \, \delta \b{u} = \b{0} \text{ on } \partial \mathcal{V}_D \biggr\} \, , \\
Q &= \biggl\{  \gamma^\alpha~:~ \gamma^\alpha, \, \dfracp{ \gamma^\alpha}{x_i} \in L^2(\mathsf{V}), \,  \gamma^\alpha = 0 \text{ on } \partial \mathcal{V}_H \biggr\} \, .
\end{align*}
Note that the slips are continuous only in the grain and are, in general, discontinuous on the grain boundary. 
For the sake of simplicity, we assume micro-free conditions on $\partial \mathcal{V}_F$.

The variational problem is obtained as follows.
The strong form of the macroscopic equilibrium equation (\ref{macro_force_balance_strong}) and the microforce balance (\ref{micro_force_balance_strong}) are tested with an arbitrary displacement $\delta \b{u} \in V$ and slip $\delta{\gamma}^\alpha \in Q$ and the result is integrated over the domain $\mathcal{V}$. 
An integration by parts is performed leading to
\begin{align}
0 &= \int_{\mathcal{V}} \b{E}(\delta \b{u}) : \b{T} ~\mathsf{d} v
- \int_{\partial \mathcal{V}_N} \delta \b{u} \cdot \b{t}^\star~\mathsf{d}a  \, ,
\label{weak_macro}\\
0 &=\int_{\mathsf{V}} \delta \gamma^\alpha \tau^\alpha ~\mathsf{d} v
-\int_{\mathsf{V}} \delta \gamma^\alpha  \pi^\alpha~\mathsf{d} v
- \int_{\mathsf{V}} \nabla \delta \gamma^\alpha \cdot \b{\xi}^\alpha~\mathsf{d} v
+  \int_{\mathsf{G}} \delta \gamma^\alpha [\b{\xi}^\alpha \cdot \b{n}]~\mathsf{d}a \label{weak_micro}
 \, .
\end{align}
The last term in \eqn{weak_micro}, arising from the integration by parts, is the weak form of the microforce balance on the grain boundary (\ref{microforce_gb}) evaluated on the interior boundaries of the grain,  can be expressed as 
\begin{align*}
 \int_{\mathsf{G}} \delta \gamma^\alpha [\b{\xi}^\alpha \cdot \b{n}]~\mathsf{d}a 
& =
\int_{\mathsf{G}} \lrb{ \delta \gamma^\alpha [\b{\xi}^\alpha_A \cdot \b{n}_A] + 
\delta \gamma^\alpha [\b{\xi}^\alpha_B \cdot \b{n}_B ]}~\mathsf{d}a \\
& =
\int_{\mathsf{G}} \lrb{ \delta \gamma^\alpha [\b{\xi}^\alpha_A \cdot \o{\b{n}}] - 
\delta \gamma^\alpha [\b{\xi}^\alpha_B \cdot \o{\b{n}} ]}~\mathsf{d}a  \\
& =
-\int_{\mathsf{G}} \lrb{ \delta \gamma^\alpha \o{\pi}{}^\alpha_A +
\delta \gamma^\alpha \o{\pi}{}^\alpha_B}~\mathsf{d}a \\
& =
-\int_{\mathsf{G}} \lrb{ \delta \gamma^\alpha 
[\o{\pi}{}^\alpha_{A,\text{en}} + \o{\pi}{}^\alpha_{A,\text{dis}}] +
\delta \gamma^\alpha 
[\o{\pi}{}^\alpha_{B,\text{en}} + \o{\pi}{}^\alpha_{B,\text{dis}}]
}~\mathsf{d}a \,.
\end{align*}

Using the above relation, the constitutive relations for the stress $\b{T}$ (\ref{elastic_constitutive}) and the (energetic) vectorial microforce (\ref{vec_microforce_constitutive}), the definition of the resolved shear stress, and the plastic flow relations in the bulk (\ref{plastic_flow_bulk}) and on the grain boundary (\ref{plastic_flow_grain_boundary}) yields the variational problem: given the surface traction $\b{t}^\star$, find the displacement $\b{u} \in V$ and the plastic slips $\gamma^\alpha \in Q$ that satisfy
\begin{gather}
0 = \int_{\mathcal{V}} \b{E}(\delta \b{u}) : \underbrace{\lrb{\dfracp{\Psi^\text{e}}{\b{E}^\text{e}}}}_{\b{T}} ~\mathsf{d} v
- \int_{\partial \mathcal{V}_N} \delta \b{u} \cdot \b{t}^\star~\mathsf{d}a \, , \label{weak_A}
\\
\begin{split}
0 &=\int_{\mathsf{V}} \delta \gamma^\alpha \underbrace{\lrb{\dfracp{\Psi^\text{e}}{\b{E}^\text{e}} : \mathbb{S}^\alpha}}_{\tau^\alpha}~\mathsf{d} v
-\int_{\mathsf{V}} \delta \gamma^\alpha  \underbrace{\lrb{\dfracp{D_\text{vis}^\alpha}{\dot{\gamma}{}^\alpha}}}_{\pi^\alpha} ~\mathsf{d} v
- \int_{\mathsf{V}} \nabla \delta \gamma^\alpha \cdot 
\underbrace{\lrb{-\dfracp{\Psi^\text{d}(\u{\rho})}{\rho^\alpha_\vdash}\b{s}^\alpha 
+ \dfracp{\Psi^\text{d}(\u{\rho})}{\rho^\alpha_\odot} \b{l}^\alpha}}_{\b{\xi}^\alpha}
~\mathsf{d} v
\\
& \quad 
- \int_{\mathsf{G}}  \delta \gamma^\alpha \lrb{ \o{\mathbb{N}}{}^\alpha_B -
 \o{\mathbb{N}}{}^\alpha_A}:\underbrace{\lrb{\dfracp{\o{\Psi}}{\o{\b{G}}}}}_{\o{\b{M}}} ~\mathsf{d}a
- \int_{\mathsf{G}}  \delta \gamma^\alpha \lrb{ \o{\mathbb{N}}{}^\alpha_B -
 \o{\mathbb{N}}{}^\alpha_A}:\underbrace{\lrb{\dfracp{\o{D}_\text{vis}}{\dot{\o{\b{G}}}}}}_{\o{\b{K}}} ~\mathsf{d}a \, ,
\end{split} \label{weak_B}
\end{gather}
for all $\delta \b{u} \in V$ and $\delta \gamma^\alpha \in Q$, where $\b{E}^\text{e}(\b{u},\u{\gamma}) = \b{E}(\b{u}) - \b{E}^\text{p}(\u{\gamma})$, $\rho^\alpha = \rho^\alpha(\nabla\gamma^\alpha)$ and $\o{\b{G}} = \o{\b{G}}(\u{\gamma}{}_A, \u{\gamma}{}_B)$.

\subsection{The incremental problem}

The time interval of interest $0 \leq t \leq T$ is partitioned into $N$ subintervals as $0 = t_0 < t_1 < \cdots < t_N = T$, with $\Delta t = t_n - t_{n-1} = T / N$.
The value of a quantity $w$ at time $t_n$ is denoted $w_n$. 
The rate of change of a quantity is approximated using an Euler-backward difference scheme as $\dot{w} \approx \Delta w / \Delta t$.
The incremental problem is obtained by evaluating relations (\ref{weak_A}) and (\ref{weak_B}) at $t_n$ as
\begin{gather}
\begin{split}
0 &= \int_{\mathcal{V}} \b{E}(\delta \b{u}) : \dfracp{\Psi^\text{e}}{\b{E}^\text{e}}\biggr|_n ~\mathsf{d} v
- \int_{\partial \mathcal{V}_N} \delta \b{u} \cdot \b{t}^\star_n~\mathsf{d}a  \label{incremental_problem_A}
\end{split} \, , \\
\begin{split}
0 &=\int_{\mathcal{V}} \delta \gamma^\alpha \dfracp{\Psi^\text{e}}{\b{E}^\text{e}}\biggr|_n : \mathbb{S}^\alpha~\mathsf{d} v
-\int_{\mathcal{V}} \delta \gamma^\alpha  \dfracp{D_\text{vis}^\alpha}{\dot{\gamma}{}^\alpha}\biggr|_{\frac{\Delta \gamma^\alpha}{\Delta t}} ~\mathsf{d} v
- \int_{\mathcal{V}} \nabla \delta \gamma^\alpha \cdot 
\b{\xi}^\alpha_n~\mathsf{d} v
\\
& \quad 
-\int_{\mathcal{G}}  \delta \gamma^\alpha \lrb{ \o{\mathbb{N}}{}^\alpha_B -
 \o{\mathbb{N}}{}^\alpha_A}:\dfracp{\o{\Psi}}{{\o{\b{G}}}}\biggr|_{n} ~\mathsf{d}a \, 
-\int_{\mathcal{G}}  \delta \gamma^\alpha \lrb{ \o{\mathbb{N}}{}^\alpha_B -
 \o{\mathbb{N}}{}^\alpha_A}:\dfracp{\o{D}_\text{vis}}{\dot{\o{\b{G}}}}\biggr|_{\frac{\Delta \o{\b{G}}}{\Delta t}} ~\mathsf{d}a \, ,\label{incremental_problem_B}
\end{split} 
\end{gather}
where
\begin{align*}
\dfracp{D_\text{vis}^\alpha}{\dot{\gamma}{}^\alpha}\biggr|_{\frac{\Delta \gamma^\alpha}{\Delta t}} 
&= 
S \lrb{ \dfrac{\vert \Delta \gamma^\alpha \vert}{\Delta t \, \dot{\gamma}{}^\alpha_0}}^m \text{sgn} \Delta \gamma^\alpha \, , \\
\dfracp{\o{D}_\text{vis}}{\dot{\o{\b{G}}}}\biggr|_{\frac{\Delta \o{\b{G}}}{\Delta t}} 
&=
\overline{S}
	\lrb{\dfrac{\vert \Delta \o{\b{G}} \vert}{\Delta t \, \dot{\o{{G}}}_0}}^{\o{m}}
	\dfrac{\Delta \o{\b{G}}}{ \vert \Delta \o{\b{G}} \vert} \, ,
\end{align*}
and  $\Delta \o{\b{G}}$ can be expressed as
\begin{align*}
\Delta \o{\b{G}}		&=  \sum_\alpha \lrb{ \Delta \gamma^\alpha_B {\o{\mathbb{N}}{}^\alpha_B} 
		-  \Delta \gamma^\alpha_A  {\o{\mathbb{N}}{}^\alpha_A}} \, . 
\end{align*}

\subsection{The minimization problem}\label{minprob}

The functional $J(\b{u},\u{\gamma})$ is defined by
\begin{gather}
\begin{split}
J(\b{u},\u{\gamma}) &:= 
\int_{\mathcal{V}} \Psi^\text{e}(\b{E}^\text{e})~\mathsf{d} v
 + \int_{\mathsf{V}}  \Psi^\text{d}(\u{\rho})~\mathsf{d} v
	 +  \Delta t \sum_\alpha \int_{\mathcal{V}}  D^\alpha_\text{vis}(\infrac{ \Delta {\gamma}{}^\alpha }{\Delta t}) ~\mathsf{d} v
	- \int_{\partial \mathcal{V}_N} \b{t}^\star_n \cdot \b{u} \, \mathsf{d} a \\
	& \quad  
	+ \int_{\mathsf{G}}  \o{\Psi} (\o{\b{G}}) ~\mathsf{d} a	
	+ \Delta t \int_{\mathsf{G}}  \o{D}_\text{vis} (\infrac{ \Delta {\o{\b{G}}}}{\Delta t}) ~\mathsf{d} a \, .
	\end{split} \label{J}
\end{gather}

\begin{prop}
The solution to the incremental problem (\ref{incremental_problem_A})--(\ref{incremental_problem_B}) is a minimizer of the functional $J$: that is, if $\mathsf{w}_n := (\b{u}_n,\u{\gamma}{}_n)$ is kinematically admissible  and satisfies 
\begin{align}
J(\b{u}_n,\u{\gamma}{}_n) \leq J(\widetilde{\b{u}}_n,\widetilde{\u{\gamma}}{}_n) \label{J_definition}
\end{align} 
for all $\widetilde{\mathsf{w}} := (\widetilde{\b{u}}_n,\widetilde{\u{\gamma}}{}_n) \in V \times Q$, then $(\b{u}_n,\u{\gamma}{}_n)$ is a solution of the incremental problem.
\end{prop}

\begin{a_proof}
For convenience the subscript $n$ is dropped. 
An arbitrary but admissible $\o{\b{G}}$ is denoted by $\t{\b{G}} = \o{\b{G}}(\u{\t{\gamma}}{}_A, \u{\t{\gamma}}{}_B)$.
Since $\mathsf{w}$ is a minimizer of $J$ it follows from \eqn{J_definition} that
\begin{align}
J( \mathsf{w}) \leq J ( [1-\theta]\mathsf{w} + \theta \, \widetilde{\mathsf{w}}) \label{J_minimization_problem}
\end{align}
for arbitrary but admissible $\widetilde{\mathsf{w}}$ with $0 < \theta < 1$.
Set
\begin{align*}
j^\alpha(\infrac{\Delta \gamma^\alpha}{\Delta t}) = 
	\int_{\mathcal{V}} D^\alpha_\text{vis}(\infrac{\Delta \gamma^\alpha}{\Delta t})~\mathsf{d} v && \text{and} &&
\o{j}(\infrac{\Delta \o{\b{G}}}{\Delta t}) =
 	\int_{\mathsf{G}} \o{D}{}_\text{vis}(\infrac{\Delta \o{\b{G}}}{\Delta t})~\mathsf{d} a \, .
\end{align*}

From the definition (\ref{J_definition}) of the functional $J$ and the convexity of $D^\alpha_{\text{vis}}$ and $\o{D}_\text{vis}$,  \eqn{J_minimization_problem} becomes
\begin{align}
& \int_{\mathcal{V}} \Psi\lrrb{\mathsf{w}}~\mathsf{d} v 
+ \Delta t \sum_\alpha j^\alpha \lrrb{ \infrac{ \Delta \gamma^\alpha}{\Delta t}} 
 + \Delta t \o{j} \lrrb{ \infrac{\Delta \o{\b{G}}}{\Delta t} }
 \notag  \\
& \quad \leq
\int_{\mathcal{V}} \Psi \lrrb{ [1-\theta]\mathsf{w} + \theta \, \widetilde{\mathsf{w}} }~\mathsf{d} v
+ [1-\theta]\Delta t \sum_\alpha j^\alpha \lrrb{ \infrac{\Delta \gamma^\alpha}{\Delta t} }
+ \theta \Delta t \sum_\alpha j^\alpha \lrrb{\infrac{ \Delta \t{\gamma}{}^\alpha}{\Delta t} } \notag \\
& \qquad - \theta \int_{\partial \mathcal{V}_N} \b{t}^\star \cdot [\t{\b{u}} - \b{u}] ~ \mathsf{d} a 
+ [1-\theta]\Delta t\o{j} \lrrb{ \infrac{ \Delta \o{\b{G}}}{\Delta t} }
 + \theta\Delta t \o{j}\lrrb{ \infrac{\Delta \t{\b{G}} }{\Delta t}}
\, . \label{min_step_A}
\end{align} 
Rearrangement of the terms in \eqn{min_step_A} leads to the inequality
\begin{gather}
\begin{split}
&\int_{\mathcal{V}} \dfrac{1}{\theta} \lrb{ \Psi \lrrb{ \lrb{1-\theta}\mathsf{w} + \theta \, \widetilde{\mathsf{w}} } - \Psi \lrrb{ \mathsf{w} }}~\mathsf{d} v
+  \Delta t \sum_\alpha \lrb{ j^\alpha \lrrb{ \infrac{ \Delta \t{\gamma}{}^\alpha}{\Delta t} } -  j^\alpha \lrrb{ \infrac{ \Delta \gamma{}^\alpha}{\Delta t} }} \\
&\qquad - \int_{\partial \mathcal{V}_N} \b{t}^\star \cdot [\t{\b{u}} - \b{u}] ~ \mathsf{d} a
+ \Delta t \lrb{ \o{j} \lrrb{ \infrac{ \Delta \t{\b{G}}}{\Delta t} } 
-  \o{j}\lrrb{\infrac{\Delta \o{\b{G}}}{\Delta t} }}
 \geq 0 \, . \label{min_step_B}
\end{split}
\end{gather}
By letting $\theta$ go to $0$ and using the definition of the derivative, \eqn{min_step_B} becomes
\begin{equation}
\begin{split}
\int_{\mathcal{V}} \dfracp{\Psi}{\mathsf{w}} : \lrb{\t{\mathsf{w}}  - \mathsf{w}} ~\mathsf{d} v
&+ \Delta t \sum_\alpha \lrb{ j^\alpha \lrrb{\infrac{ \Delta \t{\gamma}{}^\alpha}{\Delta t}} 
-  j^\alpha \lrrb{ \infrac{\Delta \gamma{}^\alpha}{\Delta t} }} \\
&- \int_{\partial \mathcal{V}_N} \b{t}^\star \cdot [\t{\b{u}} - \b{u}] ~ \mathsf{d} a
+ \Delta t \lrb{ \o{j} \lrrb{ \infrac{\Delta \t{\b{G}}}{\Delta t} } 
-  \o{j}\lrrb{\infrac{\Delta \o{\b{G}} }{\Delta t}}} \geq 0 \, .
\end{split} \label{min_step_C}
\end{equation}
Noting that
\begin{align*}
\t{\mathsf{w}} - \mathsf{w} &= \underbrace{[\t{\mathsf{w}}  - {\mathsf{w}}_{n-1}]}_{\h{\mathsf{w}}} 
	- \underbrace{[\mathsf{w} - \mathsf{w}_{n-1}]}_{\Delta \mathsf{w}} \quad 
\implies \qquad \Delta \t{\mathsf{w}}= \h{\mathsf{w}} \, ,  
\end{align*}
\eqn{min_step_C} becomes
\begin{gather}
\begin{split}
&\int_{\mathcal{V}} \dfracp{\Psi}{\mathsf{w}} : \lrb{ \h{\mathsf{w}} - \Delta \mathsf{w}} ~\mathsf{d} v
+  \Delta t \sum_\alpha \lrb{ j^\alpha \lrrb{ \infrac{ \h{\gamma}{}^\alpha}{\Delta t}} 
-  j^\alpha \lrrb{ \infrac{\Delta \gamma{}^\alpha}{\Delta t}}} 
- \int_{\partial \mathcal{V}_N} \b{t}^\star \cdot [\h{\b{u}} - \Delta \b{u}] ~ \mathsf{d} a \\
&\qquad + \Delta t \lrb{\o{j}\lrrb{ \infrac{ \h{\b{G}}}{\Delta t} } 
-  \o{j} \lrrb{ \infrac{ \Delta \o{\b{G}}}{\Delta t}}}
 \geq 0 \, . \label{min_step_D}
\end{split}
\end{gather}
Now
\begin{align*}
\dfracp{\Psi}{\mathsf{w}} : \h{\mathsf{w}} 
&= 
\dfracp{\Psi^\text{e}}{\b{E}^\text{e}} : \h{\b{E}}{}^\text{e}
+
\sum_{\alpha,\beta} \dfracp{\Psi^\text{d}}{\rho^\beta}\dfracp{\rho^\beta}{\nabla \gamma^\alpha}  \cdot \nabla \h{\gamma}{}^\alpha  \notag \\
&= 
\b{T} : \h{\b{E}}{}^\text{e}
+
\sum_\alpha \b{\xi}^\alpha \cdot \nabla \h{\gamma}{}^\alpha \, . 
\end{align*}
Hence, using the additive decomposition of $\b{E}$, \eqn{min_step_D} becomes 
\begin{gather}
\begin{split}
0 & \leq \int_{\mathcal{V}} \b{T} : \biggl[ \b{E}(\h{\b{u}}) - \b{E}(\Delta \b{u}) 
	- \sum_\alpha\lrb{ \h{\gamma}{}^\alpha - \Delta \gamma^\alpha}\mathbb{S}^\alpha \biggr]~\mathsf{d} v
	+ \int_{\mathcal{V}} \sum_\alpha \b{\xi}^\alpha \cdot \nabla ( \h{\gamma}{}^\alpha - \Delta \gamma^\alpha )~\mathsf{d} v \\
& \quad + \Delta t \sum_\alpha \lrb{ 
j^\alpha \lrrb{ \infrac{ \h{\gamma}{}^\alpha}{\Delta t}} 
-  j^\alpha \lrrb{ \infrac{\Delta \gamma{}^\alpha}{\Delta t} }} 
- \int_{\partial \mathcal{V}_N} \b{t}^\star \cdot [\h{\b{u}} - \Delta \b{u}] ~ \mathsf{d} a
+ \Delta t \lrb{\o{j}\lrrb{ \infrac{ \h{\b{G}}}{\Delta t}} 
- \o{j} \lrrb{ \infrac{\Delta \o{\b{G}} }{\Delta t}}
} \, . \label{min_step_F}
\end{split}
\end{gather}
Now set $\h{\gamma}{}^\alpha \equiv \Delta \gamma^\alpha$, which implies $\h{\b{G}} \equiv \Delta \o{\b{G}}$, to get
\begin{align*}
\int_{\mathcal{V}} \b{T} : \lrb{ \b{E}(\h{\b{u}}) - \b{E}(\Delta \b{u})}~\mathsf{d} v
 - \int_{\partial \mathcal{V}_N} \b{t}^\star \cdot \lrb{\h{\b{u}} - \Delta \b{u}} ~ \mathsf{d} a \geq 0 \, .
\end{align*}
Choose $\h{\b{u}} = \pm \Delta \b{u}$ to get
\begin{align*}
\int_{\mathcal{V}} \b{T} : \b{E}(\h{\b{u}}) ~\mathsf{d} v
 = 
 \int_{\partial \mathcal{V}_N} \b{t}^\star \cdot \h{\b{u}} ~ \mathsf{d} a  \, ,
\end{align*}
which is (\ref{incremental_problem_A}) in the incremental problem.
This leaves in \eqn{min_step_F} the inequality 
\begin{align*}
& -\int_{\mathcal{V}} \b{T} : \mathbb{S}^\alpha 
	\sum_\alpha\lrb{ \h{\gamma}{}^\alpha - \Delta \gamma^\alpha}~\mathsf{d} v
+ \int_{\mathcal{V}} \sum_\alpha \b{\xi}^\alpha \cdot \nabla \lrrb{ \h{\gamma}{}^\alpha - \Delta \gamma^\alpha }~\mathsf{d} v \\
& \qquad +  \Delta t \sum_\alpha \lrb{ j^\alpha \lrrb{ \infrac{\h{\gamma}{}^\alpha}{\Delta t}} 
- j^\alpha \lrrb{ \infrac{\Delta \gamma{}^\alpha}{\Delta t}}} 
+ \Delta t \lrb{\o{j} \lrrb{ \infrac{\h{\b{G}}}{\Delta t} } 
-  \o{j}\lrrb{ \infrac{\Delta \o{\b{G}}}{\Delta t} }} \geq 0 \, .
\end{align*}
Choose $\h{\gamma}{}^\alpha \equiv [1 - \theta]\Delta \gamma^\alpha + \theta \, \h{\gamma}{}^\alpha$ with $0 < \theta <  1$ to obtain
\begin{align*}
&-\int_{\mathcal{V}} \sum_\alpha \tau^\alpha \theta \lrb{ \h{\gamma}{}^\alpha - \Delta \gamma^\alpha}~\mathsf{d} v
+ \int_{\mathcal{V}} \sum_\alpha \b{\xi}^\alpha \cdot \nabla \lrrb{ \theta \lrb{ \h{\gamma}{}^\alpha - \Delta \gamma^\alpha} }~\mathsf{d} v \\
& \qquad 
+  \Delta t \sum_\alpha \lrb{ 
j^\alpha \lrrb{ [1-\theta] \infrac{\Delta \gamma^\alpha}{\Delta t} 
- \theta   \infrac{\h{\gamma}{}^\alpha}{\Delta t}} 
-  j^\alpha \lrrb{ \infrac{\Delta \gamma{}^\alpha }{\Delta t}}}  \\
& \qquad + \Delta t \lrb{
\o{j} \lrrb{ [1-\theta]\infrac{\Delta \o{\b{G}}}{\Delta t} - \theta \infrac{\h{\b{G}}}{\Delta t}}
 -  \o{j} \lrrb{ \infrac{\Delta \o{\b{G}}}{\Delta t} }} \geq 0 \, .
\end{align*}
Dividing through by $\theta$ and letting $\theta$ go to zero
\begin{align*}
& -\int_{\mathcal{V}} \sum_\alpha \tau^\alpha  \lrb{ \h{\gamma}{}^\alpha - \Delta \gamma^\alpha}~\mathsf{d} v
+ \int_{\mathcal{V}} \sum_\alpha \b{\xi}^\alpha \cdot \nabla \lrrb{ \h{\gamma}{}^\alpha - \Delta \gamma^\alpha} ~\mathsf{d} v \\
 & \qquad +  \sum_\alpha \dfracp{j^\alpha}{\dot{\gamma}{}^\alpha} \biggr|_{\frac{\Delta \gamma^\alpha}{\Delta t}} \lrb{ \h{\gamma}{}^\alpha - \Delta \gamma^\alpha}
 + \dfracp{\o{j}}{\dot{\o{\b{G}}}} \biggr|_{\frac{\Delta \o{\b{G}}}{\Delta t}} : \lrb{ \h{\b{G}} - \Delta \o{\b{G}}}
  \geq 0 \, .
\end{align*}
Choose $\h{\gamma}{}^\alpha = \pm \h{\gamma}{}^\alpha + \Delta \gamma^\alpha$ 
\begin{align}
-\int_{\mathcal{V}} \sum_\alpha \tau^\alpha \, \h{\gamma}{}^\alpha~\mathsf{d} v
+ \int_{\mathcal{V}} \sum_\alpha \b{\xi}^\alpha \cdot \nabla  \h{\gamma}{}^\alpha~\mathsf{d} v 
 +  \sum_\alpha \dfracp{j^\alpha}{\dot{\gamma}{}^\alpha} \biggr|_{\frac{\Delta \gamma^\alpha}{\Delta t}}  \h{\gamma}{}^\alpha 
  + \dfracp{\o{j}}{\dot{\o{\b{G}}}} \biggr|_{\frac{\Delta \o{\b{G}}}{\Delta t}} :  \h{\b{G}}
   = 0 \, . \label{min_step_G}
\end{align}
The final two terms can be expanded as
\begin{gather*}
\begin{split}
\sum_\alpha \dfracp{j^\alpha}{\dot{\gamma}{}^\alpha} \biggr|_{\frac{\Delta \gamma^\alpha}{\Delta t}}  \h{\gamma}{}^\alpha 
&= 
\sum_\alpha \int_{\mathcal{V}}  \dfracp{D^\alpha_\text{vis}}{\dot{\gamma}{}^\alpha}\biggr|_{\frac{\Delta \gamma^\alpha}{\Delta t}}  \h{\gamma}{}^\alpha~\mathsf{d} v \, ,
\end{split} \\
\begin{split}
\dfracp{\o{j}}{\dot{\o{\b{G}}}} \biggr|_{\frac{\Delta \o{\b{G}}}{\Delta t}} : \h{\b{G}} &= 
\int_{\mathcal{G}} 
\dfracp{\o{D}_\text{vis}}{\dot{\o{\b{G}}}} \biggr|_{\frac{\Delta \o{\b{G}}}{\Delta t}} : \h{\b{G}} ~\mathsf{d} a \, .
\end{split}
\end{gather*}

Finally, substituting this result into \eqn{min_step_G} gives 
\begin{align}
\int_{\mathcal{V}} \sum_\alpha \tau^\alpha \, \h{\gamma}{}^\alpha~\mathsf{d} v
- \int_{\mathcal{V}} \sum_\alpha \b{\xi}^\alpha \cdot \nabla  \h{\gamma}{}^\alpha~\mathsf{d} v 
-
\sum_\alpha \int_{\mathcal{V}}  \pi^\alpha \h{\gamma}{}^\alpha ~\mathsf{d} v
-
\sum_\alpha \int_{\mathcal{G}} \o{\b{K}} : \lrb{ \o{\mathbb{N}}{}^\alpha_B 
		-  {\o{\mathbb{N}}{}^\alpha_A}} \h{\gamma}{}^\alpha ~\mathsf{d} a  = 0 \, . \tag*{\text{\qed}} \label{min_step_final}
\end{align}
\end{a_proof}

\noindent {\sc Remark.}\ \  The variational problem and corresponding minimization formulation are easily modified to account for the alternative form for the dissipation and flow relation presented in Sections \ref{sec:mod_G} -- \ref{sec:example_modG}. The only terms affected are those involving integrals over the grain boundary; thus, the last term in (\ref{weak_B}), repeated here for convenience, is
\begin{equation*}
- \int_{\mathsf{G}}  \delta \gamma^\alpha \lrb{ \o{\mathbb{N}}{}^\alpha_B -
 \o{\mathbb{N}}{}^\alpha_A}:\underbrace{\lrb{\dfracp{\o{D}_\text{vis}}{\dot{\o{\b{G}}}}}}_{\o{\b{K}}} ~\mathsf{d}a \, ,
\end{equation*}
and this term becomes, using the definitions in Section \ref{sec:dissfnG},
\begin{equation*}
- \int_{\mathsf{G}}  \delta \gamma^\alpha \sum_\beta\lrb{ \o{\mathbb{N}}{}^\alpha_B:  \sqrt{R^{\beta\alpha}}\widehat{\b{K}}{}^{\beta\alpha}  -
 \o{\mathbb{N}}{}^\alpha_A:\sqrt{R^{\alpha\beta}}\widehat{\b{K}}{}^{\alpha\beta}}
 ~\mathsf{d}a \, .
\end{equation*}
Likewise, in the functional (\ref{J}) for the minimization problem, the term
\begin{equation*}
	+ \Delta t \int_{\mathsf{G}}  \o{D}_\text{vis} (\infrac{ \Delta {\o{\b{G}}}}{\Delta t}) ~\mathsf{d} a \, .
\end{equation*}
is replaced by 
\begin{equation*}
		+ \Delta t \int_{\mathsf{G}}  \widehat{\cal D}(\infrac{\Delta \widehat{\b{G}}{}^{\alpha\beta}}{\Delta t}) ~\mathsf{d} a \,.
\end{equation*}
 
\section{Finite element approximation}\label{sec_fem}

The spatially discrete form of the minimization problem (\ref{J}) is solved approximately using the finite element method in conjunction with a Newton--Raphson procedure. 
The software AceGen \citep{Korelc2002} is used to describe the finite element interpolation,  and to compute the residual and (algorithmically consistent) tangent contributions directly from the prescribed functional (\ref{J}) using automatic differentiation, at the level of the quadrature point. 
This approach ensures quadratic convergence of the algorithm and greatly simplifies the implementation. 

The bulk material is discretized using 8-noded hexahedral elements. 
The nodal degrees of freedom are $(\b{u}, \u{\gamma})$.
Elements containing a grain boundary are discretized using the same element as in the bulk, but with the dimension in the direction of the grain-boundary normal far less than the other dimensions. 
Importantly however, all quantities on the grain boundary are computed and integration is performed with respect to the mid-plane of the grain-boundary element. 
Thus the grain boundary is treated as a two dimensional manifold embedded in  three-dimensional space.  
The kinematic coherence constraint (\ref{coherence}) is enforced by adding the following penalty term  $\Psi^\text{co}$ to the functional $J$:
\begin{align*}
	\Psi^\text{co} := \sfrac{1}{2} k^\text{co} \vert \jmp{\b{u}} \vert^2 \, ,
\end{align*}
where $k^\text{co} > 0$ is the penalty parameter.  
An alternative approach would be to double the number of slip degrees of freedom at the face of an element that forms the grain boundary. 
Given that a grain boundary is only 1--2 atoms thick it would be preferable to use this approach. 
However, the approach is inconvenient to implement within the particular finite element framework used here. 

The finite thickness of the grain boundary has a negligible influence on the numerical results. 
The contributions from an element containing  a grain boundary to the balance of linear momentum in the bulk is ignored. 
This is a reasonable assumption provided the thickness of the grain boundary element is chosen to be sufficiently small. 
Additionally an extension to account for possible opening and sliding \citep[see e.g.][]{Gurtin2008a} is possible within this framework.

\section{Numerical examples}\label{sec_numerical_examples}

Three example problems are presented to illustrate the key features of the grain-boundary model. 
The first example illustrates the influence of misorientation of the crystal lattice between adjacent grains in a bi-crystal. 
The second investigates the influence of the orientation of the grain boundary in a bi-crystal.
The third example is that of a face-centered-cubic polycrystal subject to tensile loading.  

The constitutive parameters, chosen where possible to match those in \citet{Beers2013}, are summarised in Table~\ref{tab_material_props}. 
The thickness of the grain-boundary elements is fixed at \si{1e-5}  \si{\micro\meter}.
Energetic contributions at the grain boundary are not accounted for in the numerical examples.

\setlength{\extrarowheight}{2pt}
\begin{table}[htb]
\caption{Constitutive parameters used for the numerical examples unless stated otherwise. 
When a parameter has been varied, the alternative value is given in braces.}
\begin{center}
\begin{tabular}{|l|l|l|l|}
\hline
First Lam\'{e} parameter 	& $\lambda$ 	& \num{1.05e-1} 	& \si{N/\micro\meter^2}   \\
Second Lam\'{e} parameter	& $\mu$ 		& \num{5.4e-2} 	& \si{N/\micro\meter^2}   \\
Burgers vector length		& $b$		& \num{2.5e-4} & \si{\micro\meter}			\\
Back stress cut-off radius 	& $R$		& \num{2e1} & \si{\micro\meter}			\\
\multirow{2}{*}{Reference slip rate}
 			& $\dot{\gamma}_0$	& \num{1} & \si{/s}	\\
 			& $\dot{\o{G}}_0$	& \num{1} & \si{/s}	\\
\multirow{2}{*}{Slip resistance}
 			& $S$		&  \num{1e-2}	& \si{N/\micro\meter^2} \\
 			& $\o{S}$	& \num{1e-1} (\num{1e1})	& \si{N/\micro\meter^2}\\
Rate sensitivity				& $m$, $\o{m}$		& \num{1}	& \\
\hline
Penalty parameter & $k^\text{co}$ & \num{1e7} &\\
\hline
\end{tabular}
\end{center}
\label{tab_material_props}
\end{table}

\subsection{Misorientation of crystal lattice between adjacent grains}\label{sec_misorientation}

\begin{figure}[!ht]
 \centering
 \includegraphics[width = \textwidth]{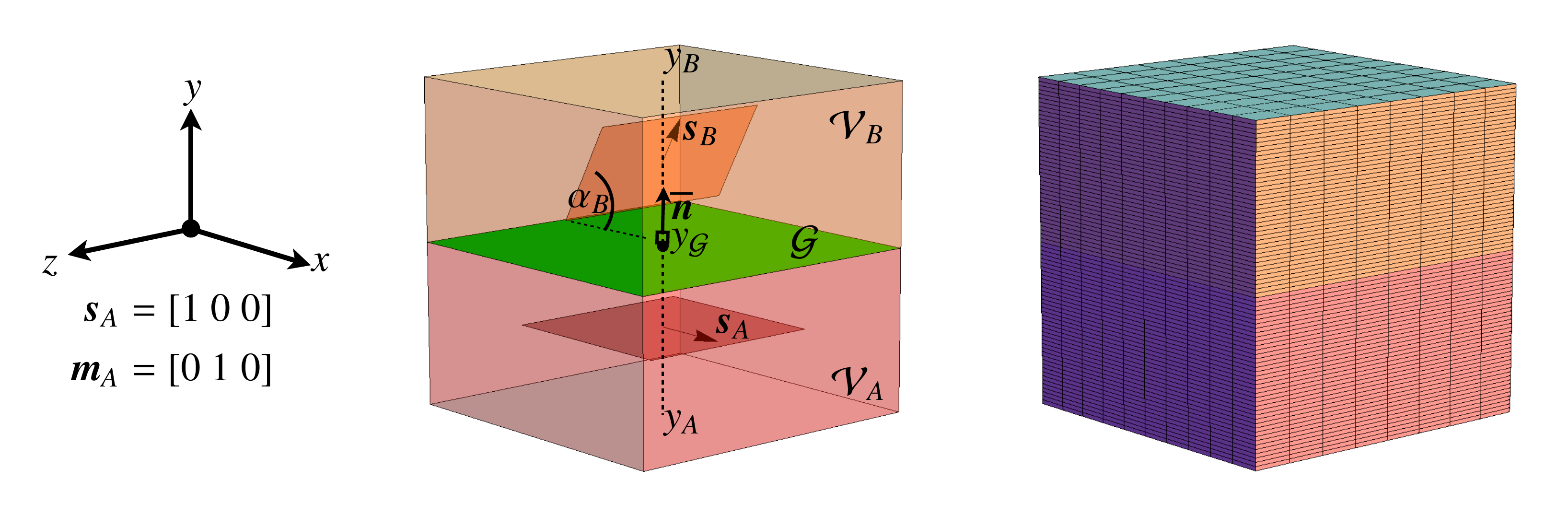}
 \caption{The domain of the misorientation problem and the computational mesh.} 
 \label{B_orientation_setup}
\end{figure}

Consider the $[50]^3$ \si{\micro\meter} bi-crystal subject to shear-type loading shown in \fig{B_orientation_setup}. 
Single slip is assumed in both grains A (lower) and B (upper). 
The crystal lattice in grain A is fixed at $(\b{s},\b{m}) = (\b{e}_1, \b{e}_2)$. 
The normal to the grain boundary is fixed at $\o{\b{n}} = \b{e}_2$. 
The crystal lattice in grain B is initially chosen as $(\b{s},\b{m}) = (\b{e}_1, \b{e}_2)$ and then rotated in increments through \ang{90} about the $\b{e}_3$ axis.
Note, the problem is similar to the one discussed in \sect{sec_comp_gurtin_beers}.
The size of the bi-crystal is small relative to the internal length scale of the gradient-plasticity model governing the bulk. 
Gradient terms will thus play a role.

Each grain is discretized using $10\times 40 \times 8$ elements. 
The lower face of grain A is prevented from displacing in the $\b{e}_2$ direction. 
The lines formed from the intersection of the lower face and the left and, separately, the back face are prevented from displacing in the $\b{e}_1$ and $\b{e}_3$ directions, respectively. 
The upper face is translated 5 \si{\micro\meter} in the $\b{e}_1$ direction. 
In addition, micro-hard boundary conditions, required for the well-posedness of the variational problem, are imposed on the upper and lower faces.

\begin{figure}[!ht]
 \centering
 \includegraphics[width = \textwidth]{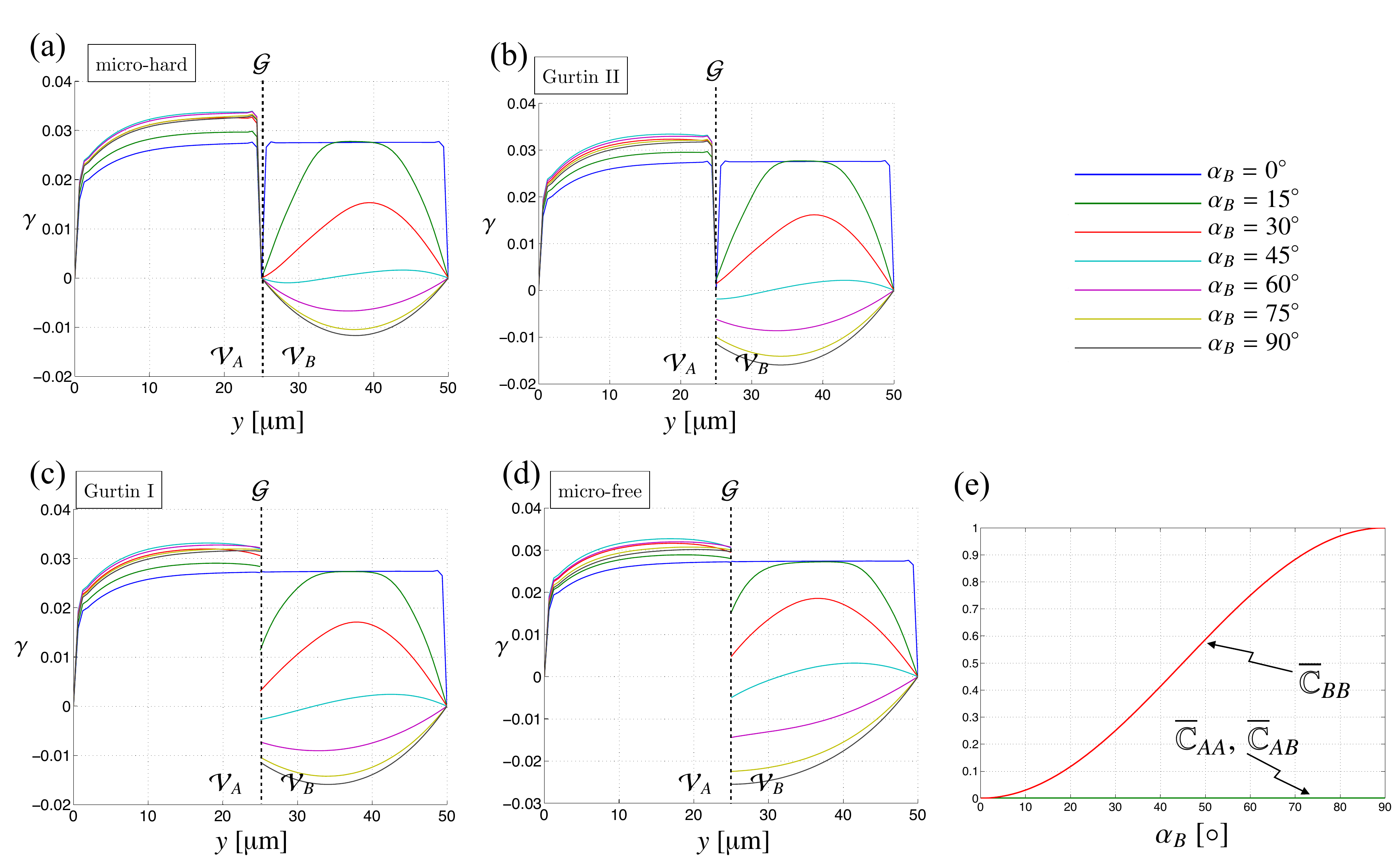}
 \caption{Distribution of $\gamma$ over the line $y_A$-$y_B$ for various choices of $\alpha_B$ in the misorientation problem for the different grain-boundary models, (a)--(d), with $\o{S}$ = \num{1e-1}.
 The interaction moduli for the range of orientations of slip system B is shown in (e).} 
 \label{orientation_gamma_1}
\end{figure}

The various interaction moduli for the range of orientations of slip system B, shown in  \fig{orientation_gamma_1}~(e), can be extracted from \fig{gb_interaction} by setting $\alpha_\mathcal{G} = 0$.
The interaction moduli $\o{\mathbb{C}}_{AA}$ and $\o{\mathbb{C}}_{AB}$ are zero as $\b{m}_A \times \o{\b{n}} = \b{0}$. 
The intra-action modulus $\o{\mathbb{C}}_{BB}$ varies smoothly between 0 and 1 as $\alpha_B$ is varied. 

The variation of $\gamma$ over the vertical line $y_A$-$y_B$ passing through the centre of the bi-crystal $y_\mathcal{G}$ for  $\alpha_B = $ \{\ang{0}, \ang{15}, \ang{30}, \ang{45}, \ang{60}, \ang{75}, \ang{90}\} corresponding the micro-hard, \citeauthor{Gurtin2008}~I and II, and the micro-free grain-boundary models is shown in \fig{orientation_gamma_1}. 
Due to the relatively small specimen size, a significant boundary layer is observed in the upper grain for all $\alpha_B \neq 0$. 
For $\alpha_B = 0$, no dislocation pile-up occurs at the grain boundary or the external boundary as the slip system in grain B is parallel to these surfaces. 
The slip system in grain A is fixed parallel to the lower surface and the grain boundary. 
A small boundary layer is thus present in the vicinity of the lower surface. 

The interaction moduli $\o{\mathbb{C}}_{AA} = \o{\mathbb{C}}_{AB} = 0$ result in similar distributions of slip in grain A for the \citeauthor{Gurtin2008}~I and micro-free grain-boundary models.
The micro-hard and \citeauthor{Gurtin2008}~II models produce a near identical distribution of slip in grain A. 
The two \citeauthor{Gurtin2008} models for the slip in grain A differ only in the vicinity of the grain boundary as the \citeauthor{Gurtin2008}~II model does not account for the interaction of the slip system and the grain boundary directly in the flow relation. 

\begin{figure}[!ht]
 \centering
 \includegraphics[width = 0.8\textwidth]{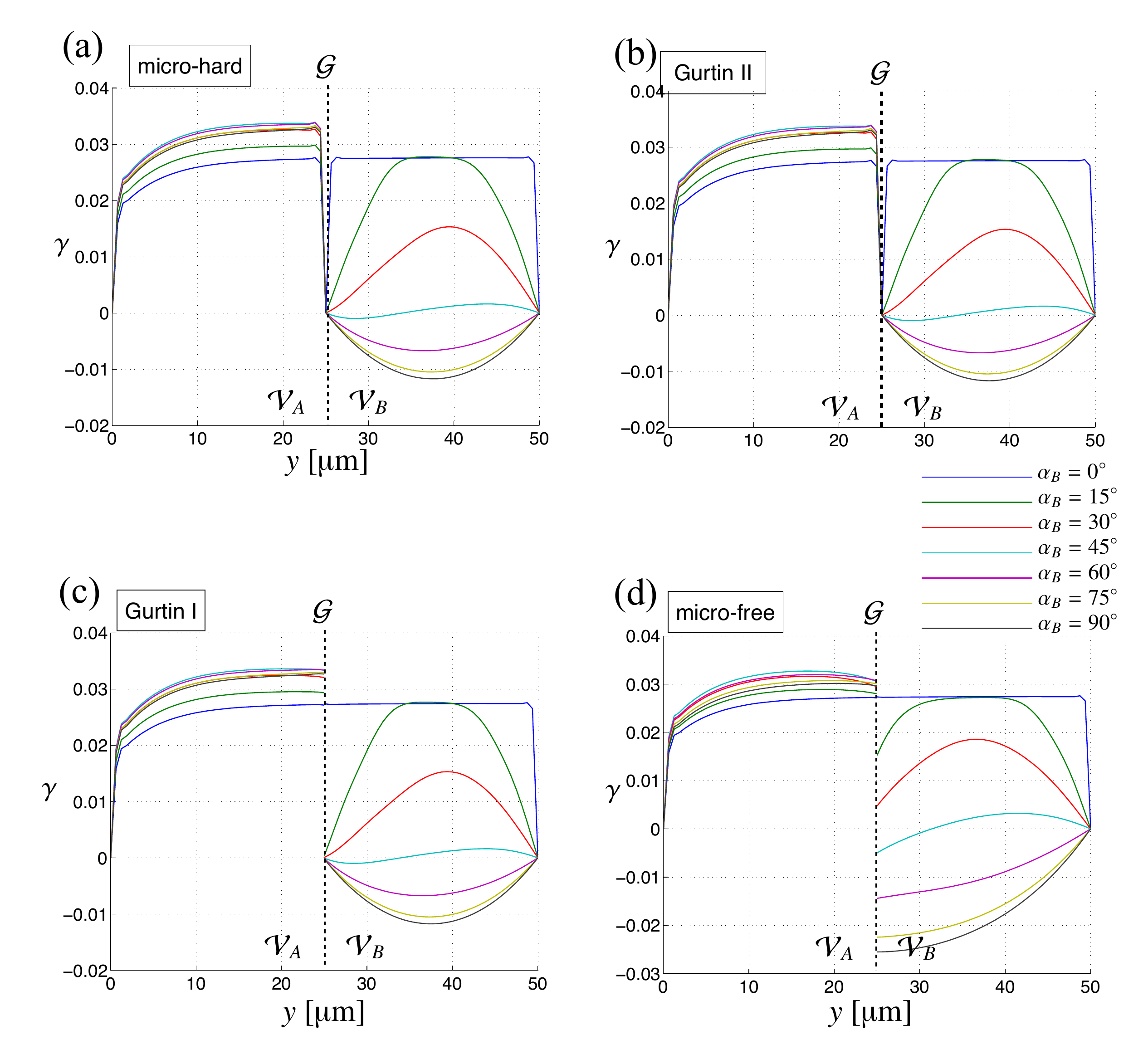}
 \caption{Distribution of $\gamma$ over the line $y_A$-$y_B$ for various choices of $\alpha_B$ in the misorientation problem for the different grain-boundary models, (a)--(d), with $\o{S}$ = \num{1e1}.} 
 \label{orientation_gamma_2}
\end{figure}

The value of the slip resistance on the grain boundary controls, in part, the flux of dislocations across the grain boundary. 
The default value of  $\o{S}$ = \num{1e-1} is relatively low and hence the flux of dislocations into grain B is relatively high.
Selecting a higher value of  $\o{S}$ = \num{1e1}, significantly decreases the flux of dislocations into the grain boundary predicted by the two \citeauthor{Gurtin2008} models, as shown in \fig{orientation_gamma_2}.

The micro-free grain boundary model is imposed by setting the grain boundary slip resistance $\o{S}$ = \num{0}. 
The micro-force balance on the grain boundary (\ref{microforce_gb}) thus reduces to the homogeneous Neumann condition.

In summary, the \citeauthor{Gurtin2008}~II model influences the flux of dislocations only through the grain boundary  slip resistance. 
The \citeauthor{Gurtin2008}~I model accounts for this and the interaction between the adjacent grains and the grain boundary. 
This observation suggests that the grain boundary  slip resistance in the  \citeauthor{Gurtin2008}~II model could be chosen as a function of the mismatch in slip system orientation at the grain boundary, thereby allowing high-angle grain boundaries to act as micro-hard.

\subsection{Orientation of the grain boundary}

\begin{figure}[!ht]
 \centering
 \includegraphics[width = 0.8\textwidth]{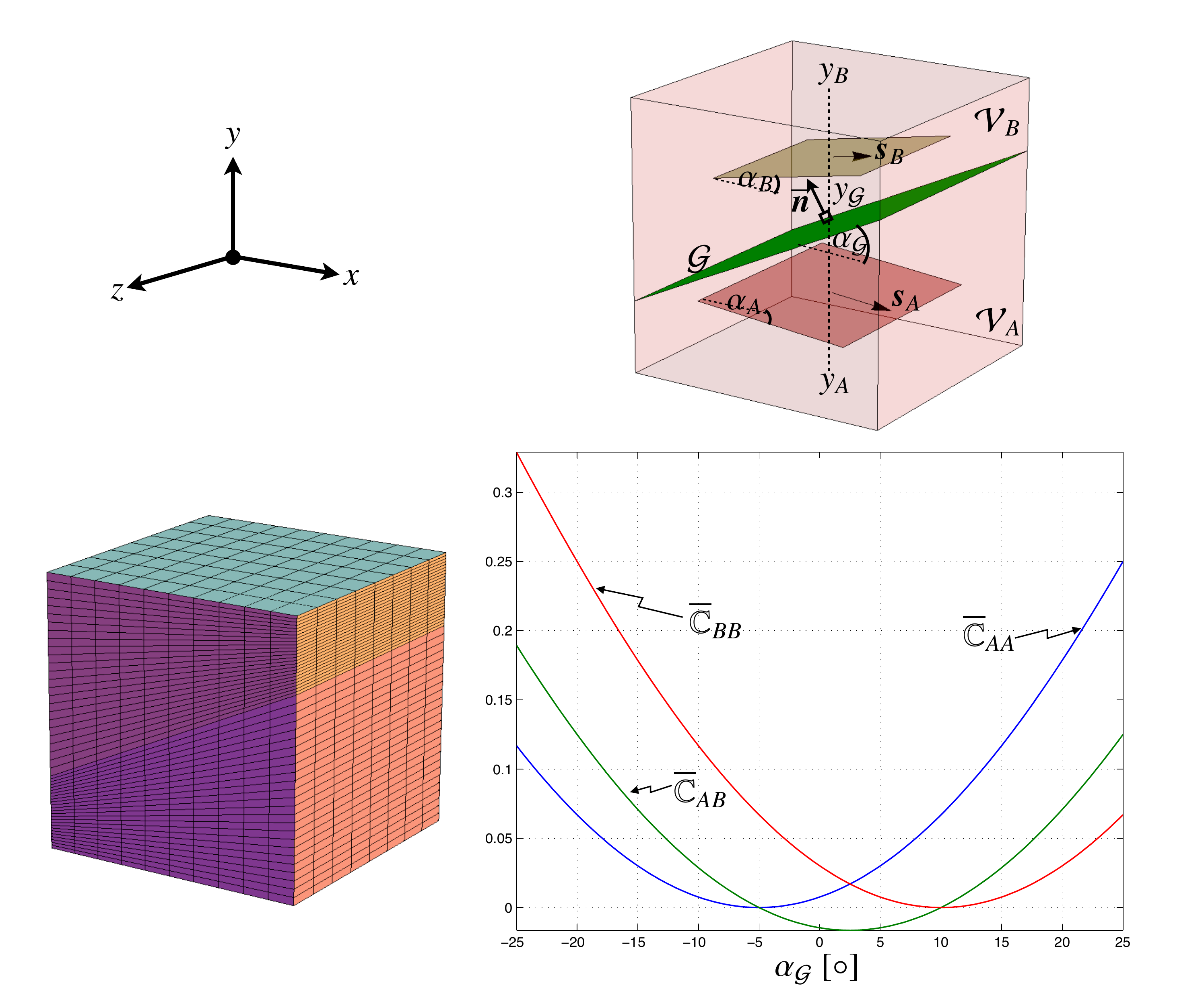}
 \caption{The domain of the grain-boundary orientation problem and the computational mesh. 
 Also shown is the variation in the inter- and intra-action moduli for a range of $\alpha_\mathcal{G}$.} 
 \label{gb_setup}
\end{figure}

The influence of the orientation of the grain-boundary is investigated by fixing the crystal lattice in both grains of the bi-crystal considered in \sect{sec_misorientation} and varying the grain-boundary orientation $\alpha_\mathcal{G}$ between \ang{-25} and \ang{25}, see \fig{gb_setup}. 
In addition, the contribution of the gradient terms in the bulk material is investigated by considering two bi-crystal sizes: a large $[l]^3=[100]^3$ \si{\micro\meter} and a smaller $[l]^3=[50]^3$ \si{\micro\meter}. 
The imposed displacement on the upper surface is $l/10$.  
The slip direction and slip plane normal for grains A and B are obtained by rotating the basis vectors $\b{e}_1$ and $\b{e}_2$ by \ang{-5} and \ang{10}, respectively. 
The variation in the inter- and intra-grain interaction moduli for the range of $\alpha_\mathcal{G}$ is given in \fig{gb_setup}. 
The angle between the normals to the slip systems and the grain boundary is relatively small for the range of $\alpha_\mathcal{G}$ used, resulting in relatively low values for the interaction moduli. 
The bi-crystal is discretized using $10\times 40 \times 8$ elements.

\begin{figure}[!ht]
 \centering
 \includegraphics[width = \textwidth]{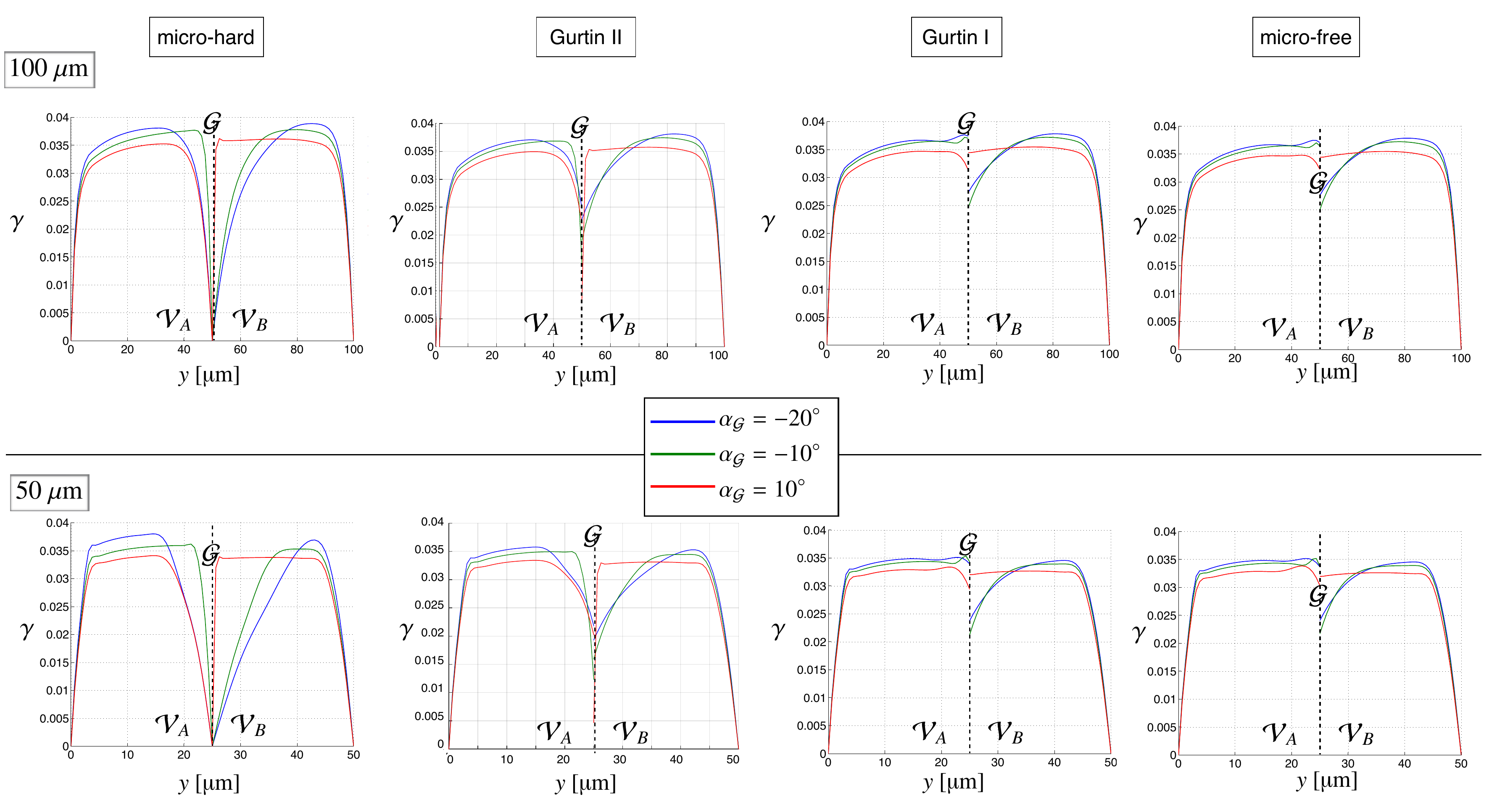}
 \caption{Distribution of $\gamma$ over the line $y_A$-$y_B$ for various choices of $\alpha_\mathcal{G}$ and specimen size in the grain-boundary orientation problem.} 
 \label{gb_gamma}
\end{figure}

The influence of specimen size for the various grain-boundary models shown in \fig{gb_gamma}. 
The boundary layers increase with decreasing size.  
The growth of the boundary layer is more marked for the micro-hard model.

For $\alpha_{\mathcal{G}}$ = \ang{10}, only $\o{\mathbb{C}}_{AA}\neq 0$, and the micro-free and \citeauthor{Gurtin2008}~I models produce near-identical results. 
As in the previous example, the relatively small value for $\o{S}$ = \num{1e-1} results in similar distributions of $\gamma$ for the micro-free and \citeauthor{Gurtin2008}~I models.
The \citeauthor{Gurtin2008}~II model behaves more like the micro-hard model. 

\subsection{Multiple slip in a polycrystal}\label{sec:multi_slip}

Consider a $[100]^3$ \si{\micro\meter} polycrystal subject to tensile loading. 
The polycrystal is composed of 27 equal sized grains.
The crystal lattice in each grain is that of a face-centered-cubic material with the orientation of the 12 slip systems relative to the unit cell shown in Table~\ref{tab_fcc}. 
The orientation of the unit cell in each grain is listed in Table~\ref{tab_euler}.

The purpose of the polycrystal example is to explore the various choices of grain boundary models for reasonably complex, three-dimensional problems. 
As seen for the single-slip examples, a micro-free grain boundary model allows for a high flux of dislocations while the micro-hard model acts as an impenetrable barrier resulting in dislocation pile-up. 
The ability of the two \citeauthor{Gurtin2008} grain boundary models to impede dislocation flow is controlled, in part, by the slip resistance of the grain boundary. 
The \citeauthor{Gurtin2008}~I model also accounts for the slip system interaction at the grain boundary. 
The ability of the two \citeauthor{Gurtin2008} models to capture the spectrum of behaviour between the micro-hard and micro-free limits is investigated by examining a range of grain boundary resistances from a low $\o{S}$ = \num{1e-4} to an extremely high value of $\o{S}$ = \num{1e10}.  
The alternative measure of the mismatch at the grain boundary developed in \sect{sec:mod_G} is also examined. 

\begin{figure}[!ht]
 \centering
 \includegraphics[width = 0.65\textwidth]{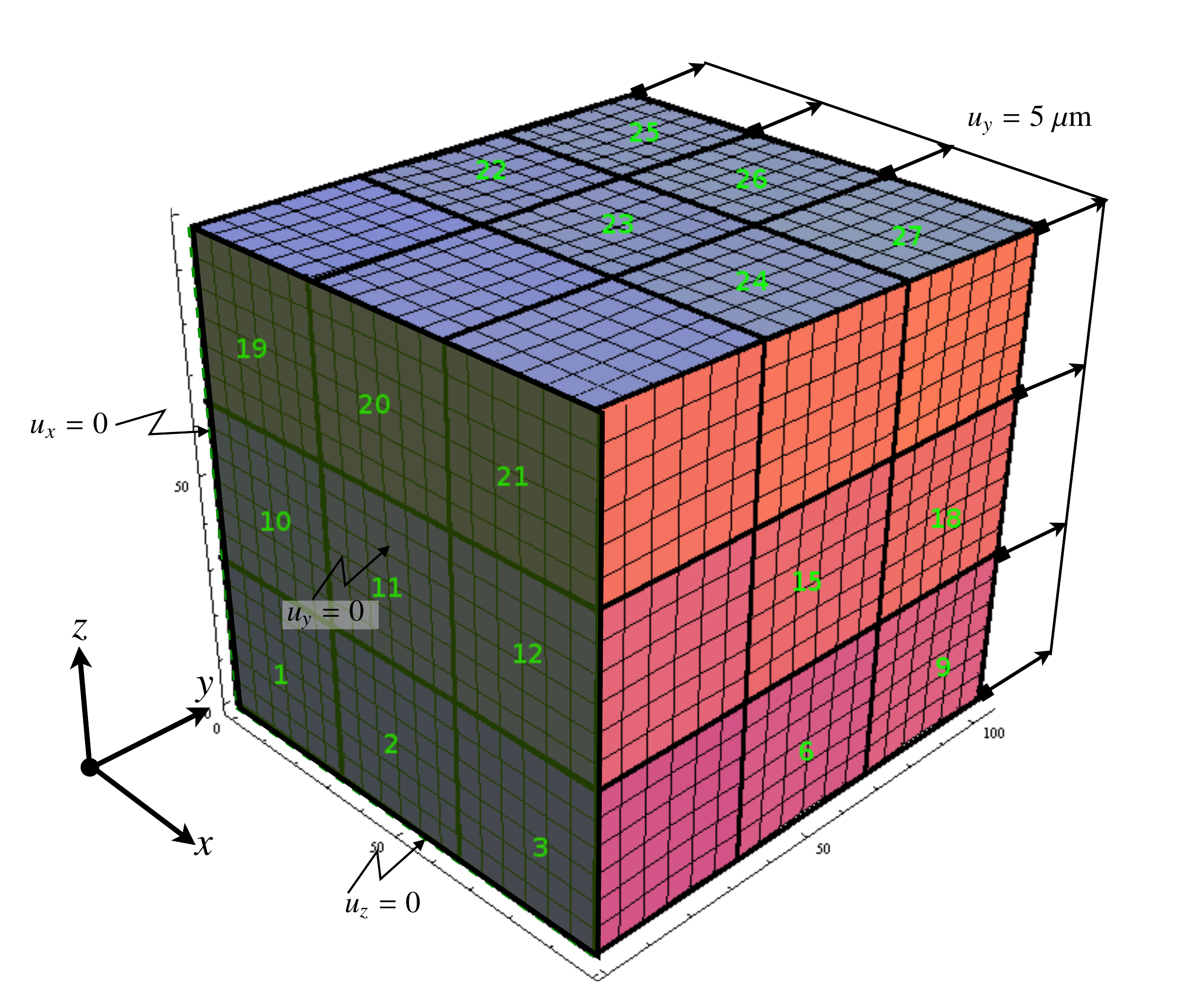}
 \caption{Discretization of the polycrsytal composed of 27 grains. 
 The numbering corresponds to the grains listed in Table~\ref{tab_euler}. } 
 \label{poly_mesh}
\end{figure}

The discretization of the domain and the boundary conditions are shown in \fig{poly_mesh}. 
Each grain is discretized with $6^3$ elements.
A displacement of $u_y=5$ \si{\micro\meter} is applied on the boundary with outward normal $\b{n} = [0,1,0]$.
The opposite boundary with outward normal $\b{n} = [0,-1,0]$ is prevented from displacing in the $y$-direction. 
The additional constraints to prevent rigid body motion are indicated in \fig{poly_mesh}.

\setlength{\extrarowheight}{2pt}
\begin{table}[htb]
\caption{The orientation of the slip planes relative to the unit cell.}
\begin{center}
\begin{tabular}{l c c l c c l c c}
\hline
$\alpha$ & $\b{s}^\alpha$ & $\b{m}^\alpha$ & $\alpha$ & $\b{s}^\alpha$ & $\b{m}^\alpha$ & $\alpha$ & $\b{s}^\alpha$ & $\b{m}^\alpha$ \\
\hline
1    & $[\o{1}~1~0]$ & $(1~1~1 )$ &
5    & $[1~0~1]$    & $(1~\o{1}~\o{1} )$ &
9    & $[0~\o{1}~\o{1}]$    & $(\o{1}~1~\o{1} )$ \\
2    & $[1~0~\o{1}]$ & $(1~1~1 )$ &
6    & $[0~1~\o{1}]$    & $(1~\o{1}~\o{1} )$ &
10    & $[1~\o{1}~0]$    & $(\o{1}~\o{1}~1 )$ \\
3    & $[0~\o{1}~1]$ & $(1~1~1 )$ &
7    & $[1~1~0]$    & $(\o{1}~1~\o{1} )$ &
11    & $[\o{1}~0~\o{1}]$    & $(\o{1}~\o{1}~1 )$ \\
4    & $[\o{1}~\o{1}~0]$ & $(1~\o{1}~\o{1} )$ &
8    & $[\o{1}~0~1]$    & $(\o{1}~1~\o{1} )$ &
12    & $[0~1~1]$    & $(\o{1}~\o{1}~1 )$ \\
\hline
\end{tabular}
\end{center}
\label{tab_fcc}
\end{table}

\begin{table}[htb]
\caption{The  Euler angles, expressed as $\mathcal{\psi}$=[$\phi_1$, $\Phi$, $\phi_2$] following the convention of \citet{Bunge1969},  for each of the grains in the polycrystal. }
\begin{center}
\begin{tabular}{l l l l l l }
\hline
Grain  & $\mathcal{\psi}$ [$^\circ$] & Grain  & $\mathcal{\psi}$ [$^\circ$] & Grain  & $\mathcal{\psi}$ [$^\circ$] \\
\hline
1& [243,166,220] & 10 & [129,160,356] & 19& [0,154,252] \\
2& [359,171,65] & 11 & [200,103,71] & 20& [239,78,296] \\
3& [260,142,153] & 12 & [29,35,196] & 21& [123,139,0] \\
4& [146,32,131] & 13 & [87,15,331] & 22& [0,85,106] \\
5& [258,7,296] & 14 & [337,133,240] & 23& [39,166,311] \\
6& [100,133,240] & 15 & [315,38,36] & 24& [40,134,327] \\
7& [328,74,214] & 16 & [97,161,67] & 25& [135,158,14] \\
8& [65,46,98] & 17 & [178,39,276] & 26& [98,99,258] \\
9& [160,86,359] & 18 & [250,152,211] & 27& [275,79,118] \\
\hline
\end{tabular}
\end{center}
\label{tab_euler}
\end{table}

\begin{figure}[!ht]
 \centering
 \includegraphics[width = \textwidth]{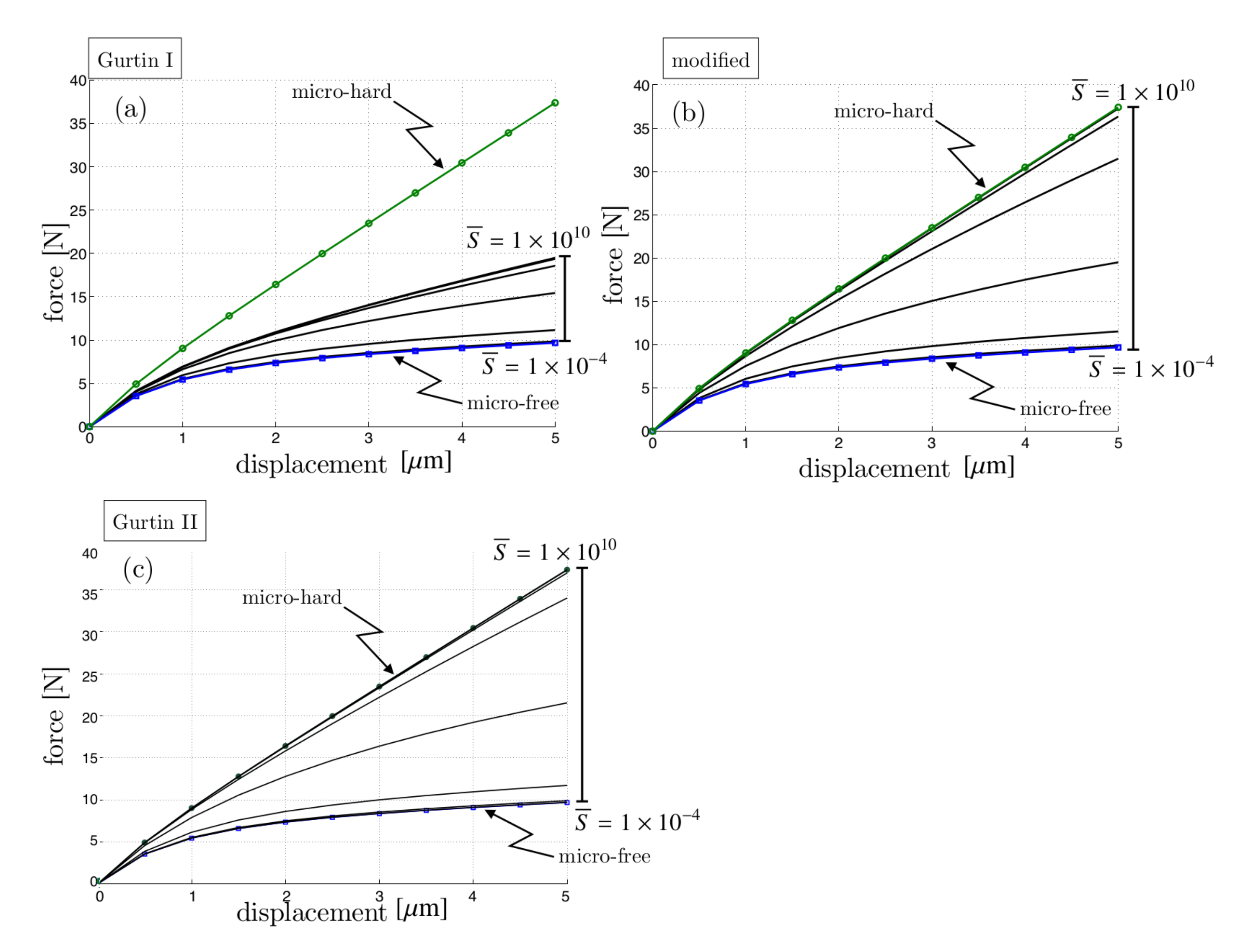}
 \caption{Applied displacement versus the  $y$-component of resultant force on the right boundary for various grain-boundary models.  
 The results of the \citeauthor{Gurtin2008}~I model are shown in (a), the modified formulation in (b), and the \citeauthor{Gurtin2008}~II model in (c).} 
 \label{poly_f_v_disp}
\end{figure}

The $y$-component of the resultant traction on the right boundary for the \citeauthor{Gurtin2008}~I, micro-hard and micro-free models is plotted against the prescribed displacement  in \fig{poly_f_v_disp}~(a).
As expected the micro-free condition provides a lower bound for the \citeauthor{Gurtin2008} models for very low values of $\o{S}$. 
From the theory and the numerical investigations involving single slip it is reasonable to expect the micro-hard condition to be an upper bound that is approached with increasing $\o{S}$. 
It's important to note that the extremely high values of $\o{S}$ chosen are not physically motivated, rather they penalize the response at the grain boundary. 
For the high-angle grain boundaries present in the current example it is physically reasonable that the models be capable of  producing a response close to micro-hard.
It is clear that this is not the case for the \citeauthor{Gurtin2008}~I model. 

\begin{figure}[!ht]
 \centering
 \includegraphics[width = \textwidth]{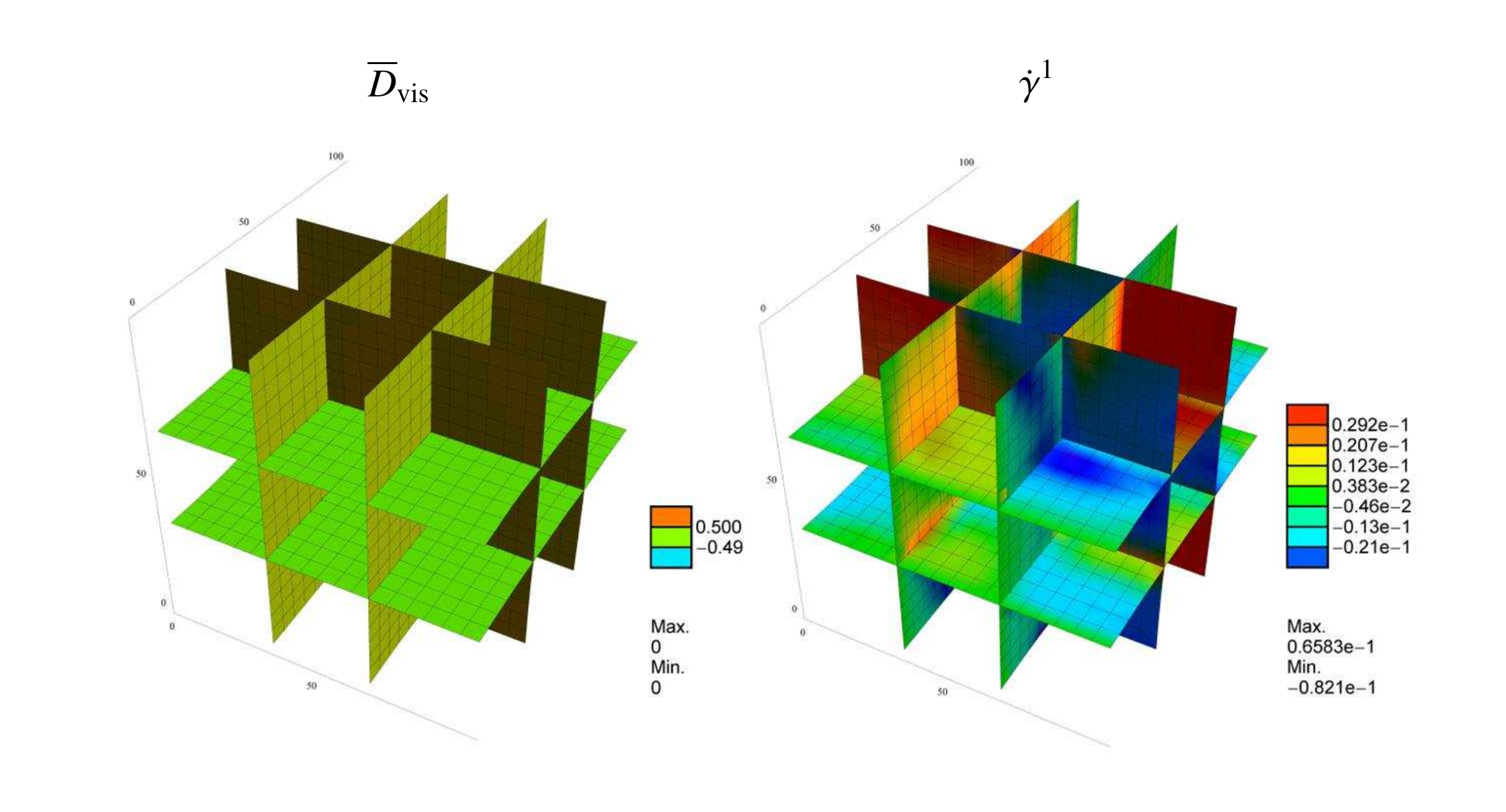}
 \caption{Distribution of $\o{D}_\text{vis}$ (see \eqn{D_vis_G}) and $\dot{\gamma}^{1}$ on each of the grain boundaries for the  \citeauthor{Gurtin2008}~I model with an artifically high value of $\o{S}$ = \num{1e10}.} 
 \label{poly_gamma_dis}
\end{figure}

Further insight into the behaviour of the \citeauthor{Gurtin2008}~I model can be obtained from the distribution of the dissipation $\o{D}_\text{vis}$, and the rate of slip on each of the grain boundaries for a high value of $\o{S}$ = \num{1e10}, as shown in \fig{poly_gamma_dis}.
As discussed in \sect{sec:grain_boundary_flow_relations}, the \citeauthor{Gurtin2008}~I model allows for a recombination of slip via the definition of the grain boundary Burgers tensor used to parametrize the flow relation. 
The numerical solution procedure discussed in \sect{sec_variational_problem} is based on the minimization of an incremental potential. 
The \citeauthor{Gurtin2008}~I model therefore permits a solution where $\o{D}_\text{vis}$ is approximately zero on the grain boundary for non-zero slip rates, as seen in  \fig{poly_gamma_dis}. 
The recombination of slip at the grain boundary in the \citeauthor{Gurtin2008}~I model is the reason that the micro-hard response can not be achieved.
It should be emphasised that this is not necessarily a limitation, but rather a feature of the model. 
This feature in the original formulation is that for non-zero slip on the grain boundary, the dissipation can be zero.
Although not shown here, the same behaviour is exhibited when energetic contributions on the grain boundary are included.

The  \citeauthor{Gurtin2008}~II model can capture the range of responses from micro-free to micro-hard, as shown in \fig{poly_f_v_disp}~(c). 
The same is true for the modified grain boundary model shown in \fig{poly_f_v_disp}~(b).

\clearpage
\section{Discussion and conclusion}\label{sec_conclusions}


The grain-boundary theory of \citet{Gurtin2008} has been compared to recently developed models in the literature. 
The theories are equivalent for planar problems.
The weak form of the governing equations has been reformulated as an incremental minimization problem. 
This reformulation provides an efficient and elegant basis for the numerical implementation. 
A series of three-dimensional numerical examples elucidated the theory for single slip in a bi-crystal. 
Various features of the \citet{Gurtin2008} grain-boundary models were illustrated for a polycrystal composed of grains with a face-centered-cubic structure.

The \citeauthor{Gurtin2008}~I model captures the geometric complexity of the grain boundary. 
A feature of the model is that it does not capture the full range of responses between micro-hard and micro-free. 
That is, it does not reproduce the widely used micro-hard limit when the grain-boundary slip resistance is used to penalize dislocation flow. 
The \citeauthor{Gurtin2008}~II model does not contain the geometrical information concerning the grain boundary but can reproduce the micro-hard limit. 
The alternative formulation based on a modified definition of the grain boundary Burgers tensor is capable of capturing the micro-hard response and contains geometrical information. 
The modified formulation reduces to the \citeauthor{Gurtin2008}~I  model for single slip. 

The computational efficiency of the formulation is greatly impacted by the choice of the grain boundary flow relation. 
The scaled calculation times relative to the micro-hard condition for the  polycrystal example are: 1.7, 4.4, 24 (for the \citeauthor{Gurtin2008}~II, \citeauthor{Gurtin2008}~I, and modified models respectively). 
The implementation of the modified model needs to be optimized, but the cost appears prohibitive. 

The numerical simulations provide valuable insight into the models.
They do not, however, allow one to judge the physical correctness of the model. 
A key challenge is therefore the validation and calibration of this and other grain-boundary models using well-devised experiments and microscopic modelling approaches (e.g.\ dislocation dynamics). 

The extension of the \citet{Gurtin2008} theory to the finite-strain regime is the subject of a companion paper in preparation. 
One significant challenge is the construction of a finite-strain counterpart to the grain-boundary Burgers tensor.
The bulk theory is well understood \citep{Gurtin2008b}. 

Elastic effects at the grain boundary have been added to the model presented here by merging it with the surface elasticity theory of \citet{Gurtin1975}. 
Results will be presented in a separate contribution.
The inclusion of thermal effects into the grain-boundary gradient-plasticity model would be a challenging and recommended extension. 
One possible extension has recently been considered by \citet{Bargmann2013}.

\section*{Acknowledgements}

DG and PW like to acknowledge the support of the German Science Foundation (Deutsche Forschungsgemeinschaft, DFG) provided for the international research training group GRK 1627.
BDR acknowledges the support provided by the National Research Foundation through the South African Research  Chair in Computational Mechanics. Part of this work was completed while BDR was visiting the Institute of Continuum Mechanics (IKM), Leibniz Universit\"at Hannover, as the recipient of a Georg Forster Research Award from the Alexander von Humboldt Foundation. The hospitality and funding provided by these organizations are acknowledged with thanks.

\appendix 

\section{Gurtin I model for model slip} \label{app_gurtin_I}

\begin{figure}[h]
\centering
\includegraphics[height=6cm,keepaspectratio=true]{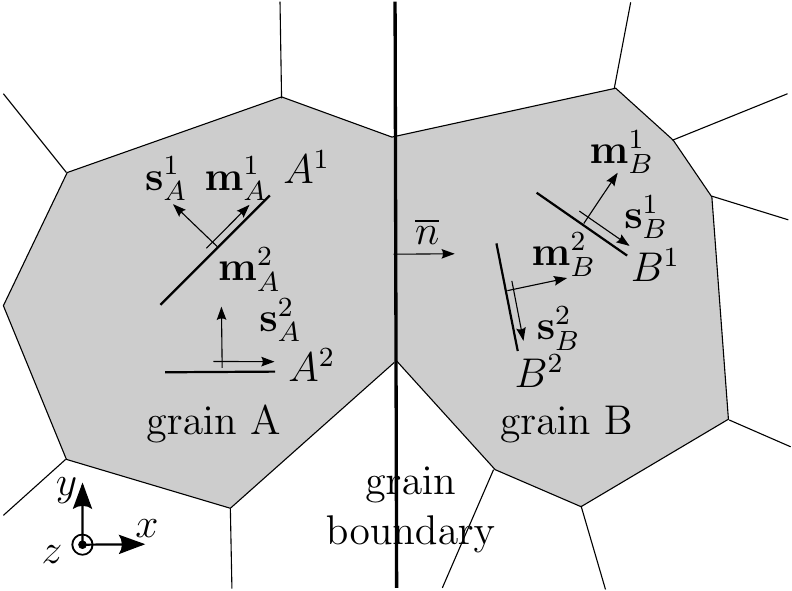}
\caption{Double slip example.  }
\label{bspdiss}
\end{figure}

The following example demonstrates how the Gurtin I model acts for multi-slip.
Recall that the dissipation function and  the rate of change of the grain-boundary Burgers tensor are given by
\begin{align*}
D_{vis}&=\frac{\overline{S}}{\overline{m}+1}\left[\frac{|\dot{\overline{\mathbf{G}}}|}{\dot{\overline{{G}}}_0}\right]^{\overline{m}+1} \dot{\overline{{G}}}_0 \, , \\
\dot{\overline{\mathbf{G}}}&=\sum_{\alpha}{[ \dot{\gamma}^\alpha_B \o{\mathbb{N}}{}_B^\alpha - \dot{\gamma}^\alpha_A \o{\mathbb{N}}{}_A^\alpha ]} \, .
\end{align*}
Assume the double-slip problem shown in  \fig{bspdiss}.

The problem is planar.
That is, all slip plane normals $\mathbf{m}_A^1,\mathbf{m}_{A}^{2},\mathbf{m}_B^1 \text{ and } \mathbf{m}_{B}^{2}$, slip directions $\mathbf{s}_A^1,\mathbf{s}_{A}^{2},\mathbf{s}_B^1 \text{ and } \mathbf{s}_{B}^{2}$ and the grain boundary normal $\o{\b{n}}$ are in the $x$--$y$-plane. 
The lattice of grain B is the lattice of grain A rotated around the z-axis. One component of $\dot{\overline{\mathbf{G}}}$ is the projection tensor $\overline{\mathbb{N}}_A^1$, that is
\begin{equation*}
\overline{\mathbb{N}}_A^1=\mathbf{s}_A^1 \otimes (\mathbf{m}_A^1 \times \mathbf{\overline{n}})=\left[ 
\begin{array}{c}
{s}_{Ax}^1 \\ {s}_{Ay}^1 \\ {s}_{Az}^1	
\end{array} \right] \otimes 
\left[
\begin{array}{c}
m_{Ay}^1 \overline{n}_z -m_{Az}^1 \overline{n}_y \\
m_{Az}^1 \overline{n}_x -m_{Ax}^1 \overline{n}_z \\
m_{Ax}^1 \overline{n}_y -m_{Ay}^1 \overline{n}_x	
\end{array}
\right]\, .
\end{equation*}
All entries in the $z$-direction are zero,  giving
\begin{equation*}
\overline{\mathbb{N}}_A^1=
\left[ 
\begin{array}{c}
s_{Ax}^1 \\ s_{Ay}^1 \\ 0	
\end{array} \right] \otimes 
\left[
\begin{array}{c}
0 \\
0 \\
m_{Ax}^1 \overline{n}_y -m_{Ay}^1 \overline{n}_x	
\end{array}
\right]=
\left[ 
\begin{array}{ccc}
0 & 0 & s_{Ax}^1 (m_{Ax}^1 \overline{n}_y -m_{Ay}^1 \overline{n}_x	) \\ 
0 & 0 & s_{Ay}^1 (m_{Ax}^1 \overline{n}_y -m_{Ay}^1 \overline{n}_x	) \\ 
0 & 0 & 0	
\end{array} \right] \, .
\end{equation*}
The term $s_{Ax}^1 (m_{Ax}^1 \overline{n}_y -m_{Ay}^1 \overline{n}_x	)$ is replaced with $a$ and $s_{Ay}^1 (m_{Ax}^1 \overline{n}_y -m_{Ay}^1 \overline{n}_x	)$ with $b$ and the same replacement is done for the other three $\overline{\mathbb{N}}_I^\alpha$. Then the rate of the Burger tensor can be written as
\begin{equation*}
\dot{\overline{\mathbf{G}}}= \dot{\gamma}_A^1 \left[ 
\begin{array}{ccc}
0 & 0 & a \\ 
0 & 0 & b \\ 
0 & 0 & 0	
\end{array} \right] + \dot{\gamma}_A^2 \left[ 
\begin{array}{ccc}
0 & 0 & c \\ 
0 & 0 & d \\ 
0 & 0 & 0	
\end{array} \right] + \dot{\gamma}_B^1 \left[ 
\begin{array}{ccc}
0 & 0 & e \\ 
0 & 0 & f \\ 
0 & 0 & 0	
\end{array} \right] + \dot{\gamma}_B^2 \left[ 
\begin{array}{ccc}
0 & 0 & g \\ 
0 & 0 & h \\ 
0 & 0 & 0	
\end{array} \right]\, .
\end{equation*}
It is clear that there are combinations of non-zero $\dot{\gamma}^\alpha_I$ leading to zero dissipation at the grain boundary. 

\section*{References}      
\bibliography{all_papers}       
\bibliographystyle{elsarticle-harv}

\end{document}